\documentclass[A4paper,12pt]{article}
\usepackage[margin=2.4cm,nohead]{geometry}
\usepackage{graphicx}
\usepackage{amsfonts}
\usepackage{amssymb}
\usepackage{amsmath}
\usepackage[english]{babel}
\usepackage{appendix}
\usepackage{lscape}
\usepackage{subcaption}
\usepackage{caption}
\usepackage{hyperref}
\usepackage{natbib}
\usepackage{setspace}
\usepackage{color} 
\newcommand{\ed}{\end{document}}
 
\begin{document}

\title{Religion and Terrorism: \\ Evidence from Ramadan Fasting\thanks{We acknowledge helpful comments by Umair Khalil, Leila Salarpour, Paul Schaudt, Austin Wright and David Yanagizawa-Drott as well as conference and seminar participants at the ASSA 2020 Annual Meeting, the CEPR Workshop on Preventing Conflicts, the CEPR Workshop on the Economics of Religion, the Albert-Ludwigs-University Freiburg, the Graduate Institute, the University of Erlangen-Nuremberg, the University of Hamburg, the University of Lausanne, and the University of St.Gallen. The authors are solely responsible for the analysis and the interpretation thereof. Roland Hodler and Paul Raschky gratefully acknowledge financial support from the Australian Research Council (ARC Discovery Grant DP150100061).}}

\author{Roland Hodler\thanks{Department of Economics, University of St.Gallen; CEPR, London; CESifo, Munich; email: roland.hodler@unisg.ch.} \and Paul A.\ Raschky\thanks{Department of Economics, Monash University; email: paul.raschky@monash.edu.} \and Anthony Strittmatter\footnote{Department of Economics, University of St.Gallen; email: anthony.strittmatter@unisg.ch.}}

\date{\today}

\maketitle

\begin{abstract}
\noindent 
We study the effect of religion and intense religious experiences on terrorism by focusing on one of the five pillars of Islam: Ramadan fasting. For identification, we exploit two facts: First, daily fasting from dawn to sunset during Ramadan is considered mandatory for most Muslims. Second, the Islamic calendar is not synchronized with the solar cycle. We find a robust negative effect of more intense Ramadan fasting on terrorist events within districts and country-years in predominantly Muslim countries. This effect seems to operate partly through decreases in public support for terrorism and the operational capabilities of terrorist groups.
\medskip

%\noindent \emph{Keywords:} Terrorism, economics of religion. \medskip

%\noindent \emph{JEL classification:} D74, H56, Z12
\end{abstract}

%%%%%%%%%%%%%%%%%%%%%%%%%%%%%%%%%%%%%%%%%%%%%%%%%%%%%%%%%%%%%%%
\newpage
\section{Introduction}\label{sec:intro}

Long before al-Qaeda's attacks on U.S. soil on September 11, 2001, terrorist groups were using religion as a justification for their actions. While many terrorist groups' motives may be of political nature, religion often serves as a common denominator among members and is used as vehicle to gain public support. Prominent examples include Abu Sayyaf in the Philippines, Hezbollah in Lebanon, the Irish Republican Army, and the Ulster Volunteer Force in Northern Ireland. Since the September 11 attacks and the subsequent U.S.-led military interventions in Afghanistan and Iraq, terrorism has been on the rise. This increase has mainly been driven by terrorist groups with links to Islam, many of which have (self-)legitimized their terrorist attacks on religious grounds. Most victims have been civilians in predominantly Muslim countries.\footnote{Evidence in support of these statements is provided in Section \ref{sec:data} and Online Appendix A.}
%\footnote{To be clear, most Muslims oppose terrorism and the ideology of Islamist groups. The Pew Research Center conducted surveys asking more than 50,000 Muslims from 13 predominantly Muslim countries (including the ten countries listed in Table \ref{tab:pew_fast} below) during the years 2002--2015 whether they feel that ``suicide bombings and other forms of violence against civilian targets are justified in order to defend Islam from its enemies.'' 78\% responded that such violence is ``never'' or ``rarely'' justified (as opposed to ``sometimes'' or ``often'').} 

This surge in terrorist attacks has made the question about the relation between religion and terrorism ever more salient. In this paper, we study the effect of intense religious experiences on terrorism,  by focusing on the daily fasting during the month of Ramadan. For our purpose, Ramadan and Ramadan fasting have three intriguing characteristics. First, Ramadan is one of the five pillars of Islam and seen as mandatory for adult Muslims until they lose their good health or sanity in older age.\footnote{People who are sick, nursing or traveling are exempt from fasting during Ramadan, but must do so later.} Table~\ref{tab:pew_fast} presents survey data by the Pew Research Center from a diverse set of ten predominantly Muslim countries.  It shows that Ramadan fasting is indeed practiced by the large majority of Muslims living in predominantly Muslim countries.
\[ \text{Table~\ref{tab:pew_fast} around here}  \]

Second, most Muslims do not only fast during Ramadan. They also engage in increased prayers and charity, abstain from sinful behavior, meet for pre-fast meals before dawn (called Suhur) and fast-breaking meals after sunset (called Iftar), and recite the Quran, which Sunni Muslims often do in extra prayers at night (called Tarawih). Hence, for most Muslims Ramadan is an intense religious experience that allows for reevaluating their lives in light of Islamic guidance as well as for socializing with family, friends and others.\footnote{Consistent with these notions, \cite{camp15} find that longer Ramadan daylight hours increase happiness despite leading to a decrease in GDP per capita; and \cite{har18} find that alimentary abstention has positive effects on pro-social behavior during Ramadan, but not when it is unrelated to religion.}
Third, Ramadan fasting lasts from dawn to sunset every day during Ramadan, and the Islamic Hijri calendar follows the lunar rather than the solar cycle. As a result, the daily fasting duration and, thereby, the intensity of the religious experience varies across locations at different latitudes in any given year, and over time at any given latitude.

While longer Ramadan hours might affect terrorism through a number of plausible channels, our theoretical argument focuses on public support for terrorism as a mechanism linking intense religious experiences to the actual occurrence of terrorism. The literature well-documents the importance of public support for the success of terrorist organizations \citep[e.g.,][]{atr03,siqu06,buen07,tes07,kru09,mal11,mal13,dav15,toft15,polo16}. Public support can include monetary assistance, in-kind assistance (e.g., weapons, vehicles or food), shelter, or legitimization. A priori, an intense religious experience may increase the perceived gap between people of the same faith and others, which could create the potential to depict others as enemies, thereby potentially increasing public support for terrorism.  Alternatively, it may induce people to see the killing of others as unacceptable, thereby decreasing public support for terrorism. 

\cite{cli09} show that participation in the Hajj pilgrimage, which 
is another intense religious experience, increases the belief in peace and harmony among adherents of different religions. In line with this literature, we document that the intense religious experience of long Ramadan fasting hours decrease the public support for terrorism in predominantly Muslim countries.
This finding is consistent with the view that most Muslims are exposed to interpretations of the Quran that condemn the use of force against innocent victims rather than interpretations preferred by Islamist groups, which focus on ``verses that have a backdrop of violence'' \citep[p.\ 19]{com17}.\footnote{The most prominent Quran verse condemning the use of force against innocent victims is verse 5:32. In its edict against terrorism in July 2005, the Fiqh Council of North America translated this verse as follows: ``Whoever kills a person [unjustly] it is as though he has killed all mankind. And whoever saves a life, it is as though he had saved all mankind.''} Our paper takes the natural next step and tests if this religiously induced change in attitude towards violence among the general Muslim population, actually leads to a reduction of violent attacks perpetrated by terrorist organizations. 
Relatedly, \cite{reese2017} argue that public support for terrorism decreases on Islamic holidays and document a lower frequency of terrorist attacks on Islamic holidays than other days. Complementary to this focus on the short-run effects of Islamic holidays, we investigate the medium-run effects of a heightened level of religious devotion and extended periods of spiritual sanctity on terrorism in the year following Ramadan. 

%OLD: Relatedly, \cite{reese2017} argue that public support for terrorism decreases on Islamic holidays and document a lower frequency of terrorist attacks on Islamic holidays. Their study estimates an overall holiday effect on terrorism by comparing the frequency of attacks between Islamic holidays and other days. In contrast, we investigate the effects of a heightened level of religious devotion and extended periods of spiritual sanctity during prolonged episodes of the Ramadan. In addition, \cite{reese2017} focus on the the short-run effects of Islamic holidays while we focus on the persistent effects of the intense religious and social experiences during Ramadan fasting.

According to the arguments made above, the decrease in public support for terrorism induced by longer fasting hours could reduce the operational capability of perpetrator groups. This could make terrorist attacks more difficult and costly, which leads us to our main hypothesis:  

\begin{center}
\begin{minipage}[c]{0.85\textwidth}
H1: Longer Ramadan fasting hours decrease terrorism in predominantly Muslim countries. 
\end{minipage}
\end{center}

However, a decrease in public support does not alter all attack types and tactics equally. Suicide attacks require less public support than tactics that require escape routes or shelter. Indeed, \cite{ber08} emphasize that suicide attacks remain effective in conditions in which other tactics would fail and are therefore chosen more often ``when conditions disfavor'' (p.\ 1948). Similarly, terrorist attacks against unarmed civilians are operationally easier and, thus, less contingent on public support than attacks against armed targets.  Assuming that the leaders of perpetrator groups are rational agents who weigh the costs and benefits of their options,\footnote{This assumption is common in theoretical studies on the organizational structure and functioning of religious groups and terrorist organizations \citep[e.g.,][]{ian92,ber08,ber11}.} our theoretical argument therefore yields a second hypothesis:

\begin{center}
\begin{minipage}[c]{0.85\textwidth}
H2: Longer Ramadan fasting hours have differential effects on terrorism in predominantly Muslim countries, with the decrease being more pronounced (i) for operationally difficult terrorist attacks than for suicide attacks, and (ii) for terrorist attacks against armed targets than for attacks against unarmed civilians.
\end{minipage}
\end{center}

In this paper, we  provide empirical evidence for these theoretical predictions.
We thereby use an identification strategy that builds on \cite{camp15}. They exploit that Ramadan daylight hours, as a measure of the daily fasting duration in Ramadan, oscillate around approximately 12 hours in any given location over the years. More specifically, the fact that the amplitudes of these oscillations are smaller at locations closer to the equator allows them to use country- and year-fixed effects to control for any country- or time-specific characteristics when estimating the effects of the daily fasting duration in Ramadan and the accompanying religious experience. 
We build on \cite{camp15} identification strategy but take it to the subnational level by exploiting variation in Ramadan daylight hours across districts within country-years, i.e., we use district- and country-year fixed effects. One reason for exploiting variation within country-years is the existing literature on the determinants of terrorism. \cite{gass11} review the literature and find 65 country-level correlates of terrorism that have been proposed in earlier studies. Moreover, the effects of many of these potential determinants vary across countries and time periods \citep[e.g.,][]{mei13,end16}. The use of country-year fixed effects implies that we implicitly control for all these potential determinants and their heterogeneous effects across countries and over time. 

To investigate our two hypotheses, we use data from the Global Terrorism Database (GTD) by the National Consortium for the Study of Terrorism and Responses to Terrorism \citep{start_data}. We build a panel dataset of administrative regions at the second subnational level (typically districts) from 190 countries and territories, and with annual frequency from 1970--2016. First, we focus on the aggregate effect to provide evidence for our main hypothesis (H1). We find that the probability of the occurrence of a terrorist event (or a deadly terrorist event) in a given district and year decreases by 2--3 percentage points in predominantly Muslim countries when the daily fasting duration in Ramadan increases by an additional hour. As one would expect, this decrease is exclusively driven by terrorist attacks by Muslim perpetrator groups. Moreover, we find no robust effect of longer Ramadan fasting on countries with a Muslim population share below 75\%.

We then look at differential effects across types and targets of terrorist attack to investigate our second hypothesis (H2). We confirm that the negative effects of longer Ramadan fasting hours on assaults and bombings are considerably stronger than the effect on suicide attacks; and that the negative effect on terrorist attacks against armed targets is larger than the effect on terrorist attacks against unarmed civilians.

Finally, we test a key element of our theoretical argument, namely the negative effect of longer Ramadan fasting hours on public support for terrorism in predominantly Muslim countries. We use survey data from 13 predominantly Muslim countries collected by the Pew Research Center. We indeed find that more intense Ramadan fasting leads to a decrease in the share of Muslim respondents who support terrorism or, more precisely, who feel that violence is often or sometimes justified ``to defend Islam from its enemies.'' %This finding holds for more and less religious people alike, which is consistent with the view that it is not only the purely religious but also the social aspects of experiences like Ramadan fasting that matters. These findings are also interesting by and in itself, as they document that religious practices have important effects on (politically relevant) ideological views.

Taken together, these three sets of results provide strong empirical support for our theoretical argument. We nevertheless also evaluate some alternative channels through which longer Ramadan daylight hours may impact on terrorism. In particular, Ramadan fasting is known to have physiological and psychological effects, such as body weight loss, sleep deprivation, tiredness, lassitude and irritability \citep[e.g.,][]{lei03}. Hence, a priori, it seems plausible that such physiological and psychological effects on (potential) terrorists or security forces could have a direct effect on terrorism. However, we show that the negative effects of longer Ramadan daylight hours on terrorist events persist beyond the first few months after Ramadan. Given that the purely physiological and psychological effects of Ramadan are unlikely to last longer than a few weeks or months, this finding makes it implausible that the physiological and psychological effects on (potential) terrorists or security forces are a main channel.\footnote{Yet another channel could be that more intense Ramadan fasting increases the empathy and self-restraint of (potential) terrorists. This idea, however, is hard to reconcile with our finding that longer Ramadan fasting hours have a comparatively smaller effect on terrorist attacks against unarmed civilians than on attacks against armed targets.}

The remainder of the paper is structured as follows: Section \ref{sec:data} describes the data, Section \ref{sec:strategy} discusses our empirical strategy, Section \ref{sec:results} presents the results, and Section \ref{sec:con} concludes.

%%%%%%%%%%%%%%%%%%%%%%%%%%%%%%%%%%%%%%%%%%%%%%%%%%%%%%%%%%%%%%%
%\newpage
\section{Data}\label{sec:data}

%%%%%%%%%%%%%%%%%%%%%%%%%%%%%%%%%%%%%%%%%%%%%%%%%%%%%%%%%%%%%%%
\subsection{Data on terrorist events}\label{sec:data.gtd}

We use the \href{https://www.start.umd.edu/gtd/}{Global Terrorism Database} (GTD) by the National Consortium for the Study of Terrorism and Responses to Terrorism \citep{start_data}, which is among the most comprehensive terrorist event datasets. The GTD defines a terrorist event as ``the threatened or actual use of illegal force and violence by a non-state actor to attain a political, economic, religious, or social goal through fear, coercion, or intimidation'' \citep[][p.\ 10]{start_codebook}. 

For an event to be included in the GTD it must fulfill the following three criteria: First, the incident must be intentional. Second, it must involve violence against humans or property, or threats thereof. Third, its perpetrators must be non-state actors. In addition, the event must meet at least two of the following three criteria: First, the perpetrators have a political, economic, religious, or social goal. Second, they intend to coerce, intimidate, or send some other message to a large audience (beyond the victims). Third, the event does not occur in the context of legitimate warfare activities \citep[][p.\ 10]{start_codebook}.

The GTD contains detailed information about the time and location of terrorist events and the number of deaths, which includes perpetrators and victims. The START assembles the GTD from three main data sources. First, the Pinkerton Global Intelligence Services identified terrorist events between 1970 and 1997 from wire services, government reports and major international newspapers. Second, the Center for Terrorism and Intelligence Studies, using archival sources, documented attacks until 2008. Third, the Institute for the Study of Violent Groups included data on terrorist events that occurred between 2008 and 2011. The START continuously integrated and updated the data sources. They improved machine learning and data mining techniques to pre-screen news articles that potentially include information about terrorist events. Currently, the GTD contains more than 180,000 terrorist events.\footnote{According to the START (2018, p.~3), the GTD does not contain any terrorist events for 1993 ``because they were lost prior to START's compilation of the GTD from multiple data collection efforts.'' Our results are robust to the exclusion of the years around 1993 (see Section \ref{sec:results_main}).} Given that the GTD combines different data sources and that the data quality may vary over time and across countries, the use of country-year fixed effects will be important.

A major advantage of the GTD is the daily temporal scale and the availability of the location of the terrorist events. This information enables us to exploit variation in those events within districts and country-years in our main analysis, and also to present additional results based on quarterly aggregates of terrorist events.

We omit terrorist events that occurred outside our sample period. % from Ramadan year 1390 to Ramadan year 1436. 
In addition, we omit 12,991 terrorist events that we cannot match to any of the 39,530 districts in our final panel dataset (see Section \ref{sec:data.panel} for details). The START only adds terrorist events from credible sources to the GTD. We drop 22,168 ambiguous events that the START classifies with high probability, but not certainty, as an act of terrorism. These ambiguous events could be insurgencies, guerrilla actions, conflicts between or within groups, or other crimes. Overall our data contains 126,346 terrorist events that caused 254,108 terror deaths. 

The GTD also provides the names of the perpetrator groups for the majority of terrorist events. We classify these perpetrator groups into Muslim and non-Muslim groups. Muslim perpetrator groups have some connection to Islam, e.g., many of their members and supporters being Muslim. 
%We further classify Muslim perpetrator groups into groups with either an Islamist or secular ideology. Islamist perpetrator groups claim to have a connection to Islam and confess to terrorist attacks citing religious ideologies or religious motives. Examples include al-Qaeda, al-Shabaab, Boko Haram and Islamic State. Secular Muslim perpetrator groups are driven by other ideologies and motives. They include separatists (or freedom fighters), such as the Kurdistan Workers' Party or the Palestine Liberation Organization, and communist groups, such as Devrimci Sol. 
For the years up to 2008, we follow the classification of \cite{kis14}. We update their classification for more recent years.
%\footnote{The classification of perpetrator groups is available upon request.} 
In total, we identify 878 Muslim and 2,180 non-Muslim perpetrator groups. These groups are jointly responsible for 51\% of terrorist events and 67\% of terror deaths in our data. We are unable to classify the perpetrators of the remaining terrorist events because these perpetrators are unknown, multiple groups confessed to the attack, or the press reports were ambiguous. 

For most terrorist events, the GTD provides information about the targets and the non-mutually exclusive attack types, which reflect the broad tactics used by the perpetrators.
The most common targets are unarmed civilians and armed targets (such as military, police or armed rivals). 
The most common attack types and the causes of most terror deaths are assaults and bombings. Below we look at suicide attacks. Most suicide attacks are assaults or bombings, but they are disproportionally deadly and have become disproportionally common in predominantly Muslim countries.

Online Appendix B provides detailed information about the terrorist events during our sample period. Table B.1 list the countries and terrorists with the highest numbers of terrorist events and terror deaths, and Table B.2 provides summary statistics stratified across country groups differing in their Muslim population shares (see Section \ref{sec:data.panel} for details on these shares). 
Figures B.1--B.5 illustrate the development of terrorist events and terror death over time -- in the aggregate as well as by country groups, types of perpetrators, attack types, and targets. These figures confirm that the rise in terrorist attacks since the September 11 attacks has mainly been driven by Muslim perpetrator groups and that most victims have been civilians in predominantly Muslim countries. In addition, Figures B.6--B.8 examine the dynamics of terrorism within years and document that terrorist events are neither particularly frequent, nor particularly infrequent during Ramadan (as compared to other months of the Islamic Hijri calender). %If terrorist events were uniformly distributed across the Islamic months, we would expect a share of 8.3\% of terrorist events during Ramadan. Figures \ref{fig:duringramadan}--\ref{fig:duringramadan_mcountries} show that these shares are indeed fluctuating around approximately 8.3\% no matter whether we look at all terrorist events or restrict attention to either terrorist events committed by Muslim perpetrator groups or terrorist events in predominantly Muslim countries. 

%%%%%%%%%%%%%%%%%%%%%%%%%%%%%%%%%%%%%%%%%%%%%%%%%%%%%%%%%%%%%%%
\subsection{Our panel dataset}\label{sec:data.panel}

Our analysis is based on panel data. The cross-sectional units are administrative regions at the second subnational level (ADM2). These units correspond to counties or, more often, districts. Compared to cross-sectional units that do not coincide with administrative regions, the use of districts has the advantage that the START typically provides the geo-references for cities or districts in which terrorist events occurred. In total, our sample contains 39,886 such districts from 190 countries and territories. 

The time-series dimension has annual frequency. However, we deviate from the common practice of defining years based on the Gregorian calender. The disadvantages of basing the analysis on Gregorian years would be two-fold: First, Ramadan would be early in some Gregorian years and late in others. Second, some Ramadans overlap two Gregorian years. In these cases, Gregorian years often contain parts of two Ramadans. 
Instead, we base our analysis on the Islamic Hijri calendar. The Islamic year has 12 months. The duration of each month is based on the lunar cycle and varies between 29 and 30 days, depending on actual observations of the lunar crescent.\footnote{Some Muslims use a tabular calendar, which assigns 30 days to odd months and 29 days to even months. In Iran and Afghanistan, the solar calendar is common, but Ramadan is practiced following the lunar Hijri calendar.} As a result, the Islamic year has 354 or 355 days. Accordingly, the Islamic year is 10--12 days shorter than the Gregorian year, which is based on the solar cycle. Consequently, the lunar cycle laps the solar cycle approximately every 33 years. 

We construct a pseudo time scale which we label ``Ramadan year.'' Ramadan is the ninth month of the Islamic year. A Ramadan year begins with Ramadan of the corresponding Islamic year and ends on the last day before Ramadan of the subsequent Islamic year. For example, Ramadan year 1390 begins on the 1st of Ramadan of the Islamic year 1390 (corresponding to the Gregorian date November 1, 1970) and ends on the 30th of Shaban, which is the eight month of the Islamic Hijri calender, in the Islamic year 1391 (corresponding to the Gregorian date October 20, 1971). %\footnote{To avoid ambiguity, we report the Gregorian year that corresponds to the beginning of the respective Ramadan year. For example, we match the Gregorian year 1970 to Ramadan year 1390.}
The reason for basing our analysis on Ramadan years is that we are interested in how the daily fasting duration during Ramadan affects terrorism over the next year. The underlying idea is that longer and more intense Ramadan fasting may shape beliefs and behavior in the entire year until the next Ramadan. We restrict our sample to Ramadan years 1390 to 1436 (corresponding to the Gregorian dates November 1, 1970, and June 5, 2016). Hence, the time-series dimension is 47 Ramadan years. Our sample therefore contains 1,874,642 observations. 

We transform the terrorism data from the GTD by aggregating the terrorist events and terror deaths by districts and Ramadan year, using the date and location information in the GTD and the administrative (ADM2) boundaries provided by GADM. In the resulting panel, the occurrence of terrorist events is highly right-skewed. There are terrorist events in only 27,866 (1.6\%) of our observations. 25,147 of these observations have between one and four terrorist events, while there were 980 terrorist events in Baghdad in Ramadan year 1436. Similarly, there are deadly terror attacks only in 17,223 (0.9\%) of our observations. 9,740 of these observations have between one and four terror deaths, but Baghdad suffered 3,348 terror deaths in Ramadan year 1427. By comparison, the terrorist attacks on the World Trade Center in New York on September 11, 2001, enter the data with 2,766 deaths. We transform our terrorism data to avoid the results being driven by extreme outliers. In the main analysis, we focus on the the extensive margin of terrorist events defined as $1\{N>0\}\cdot 100$, where $N$ stands for the number of terrorist events or terror deaths by district and Ramadan year.\footnote{We multiply our dependent variables by 100 to ease the interpretation of the coefficient estimates.} The extensive margin of terrorist events indicates whether any terrorist event occurred in the given district and Ramadan year, and the extensive margin of terror deaths indicates whether any fatal terrorist event occurred. We provide additional results for the intensive margin based on the log-modulus transformation $\ln(N+1)\cdot 100$.

In order to collect information on daylight hours during Ramadan, we first determine the centroid of each district.\footnote{More specifically, we calculate the ``representative'' centroid (point) for each ADM2 polygon, which ensures that the point falls inside the polygon.}  We then use the centroid's geo-coordinates to collect the district's daylight hours during Ramadan from the \href{http://aa.usno.navy.mil/data/docs/RS_OneDay.php}{Astronomical Application Department}. Finally, we average the daylight hours in a district over all days of Ramadan in a given Ramadan year. Hence, the Ramadan daylight hours always refer to the latest Ramadan, i.e., the first month of a Ramadan year. %This is the rationale for using the pseudo time scale ``Ramadan year'' in our analysis.

Table \ref{tab:des_panel} presents the Ramadan daylight hours and the occurrence of terrorist events and terror deaths in our panel dataset. 
\[ \text{Table \ref{tab:des_panel} around here}  \]
We thereby stratify the countries and territories into three groups based on the share of the Muslim population in these countries. We use data on the Muslim population share in 1990 from the \href{https://web.archive.org/web/20110202125115/http:/features.pewforum.org:80/muslim-population/?sort=Percent1990}{Pew Research Center}. The data are based on national censuses, demographic and health surveys, and general population surveys and studies. We distinguish between 137 predominantly non-Muslim countries and territories with a Muslim population share below 25\%, 18 religiously divided countries and territories with a share between 25\% and 75\%, and 35 predominantly Muslim countries and territories with a share above 75\%.\footnote{Online Appendix A provides more information on these countries. Table A.1 lists all countries and territories in these three country groups, and Figure A.1 shows the corresponding map. Figure A.2 shows a histogram of Muslim population shares. It shows that 70\% of the countries and territories have a Muslim population share below 5\% or above 95\%. The exact thresholds used in the stratification of the countries and territories into three groups are therefore not particularly important (see also Section \ref{sec:results_main}).}  

%%%%%%%%%%%%%%%%%%%%%%%%%%%%%%%%%%%%%%%%%%%%%%%%%%%%%%%%%%%%%%%%
%\newpage
\section{Empirical strategy}\label{sec:strategy}

Our empirical strategy builds on \cite{camp15} and exploits key characteristics of Ramadan fasting and the Islamic calender. Daily fasting from dawn to sunset during Ramadan is seen as mandatory and is practiced by many Muslims (see Table \ref{tab:pew_fast}). Daylight hours, however, vary by latitude and seasons, and the season in which Ramadan takes place varies over time because the Islamic calender is not synchronized with the solar cycle. As a result, the daily Ramadan fasting duration typically varies across districts within a given country and year, and the difference in Ramadan fasting duration between districts varies over time. 

\begin{figure}[h!]
\caption{Average daylight hours during Ramadan} \label{fig:cycle}
\centering
\begin{subfigure}{0.475\textwidth}
\includegraphics[width=\textwidth]{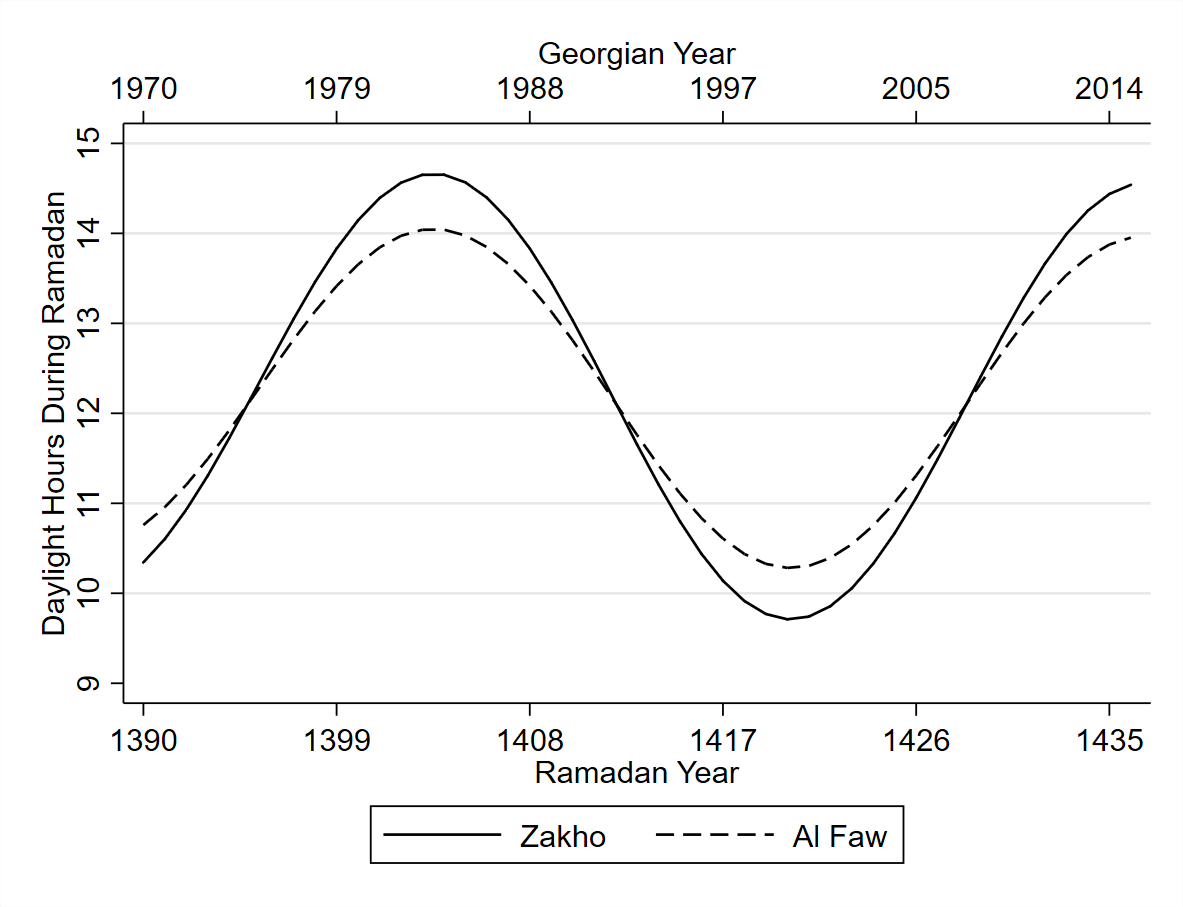}\caption{Iraq}
\end{subfigure}
\begin{subfigure}{0.475\textwidth}
\includegraphics[width=\textwidth]{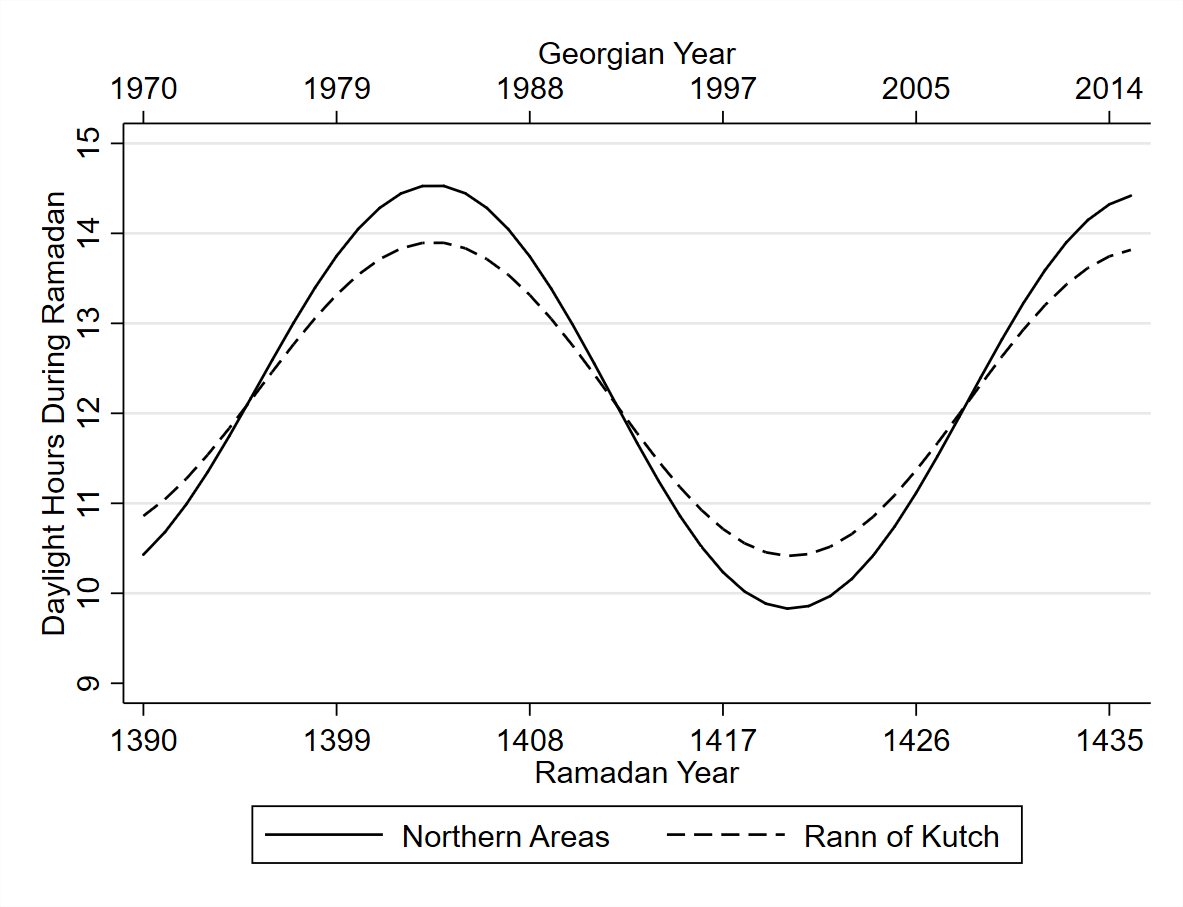}\caption{Pakistan}
\end{subfigure}
\\
\begin{subfigure}{0.475\textwidth}
\includegraphics[width=\textwidth]{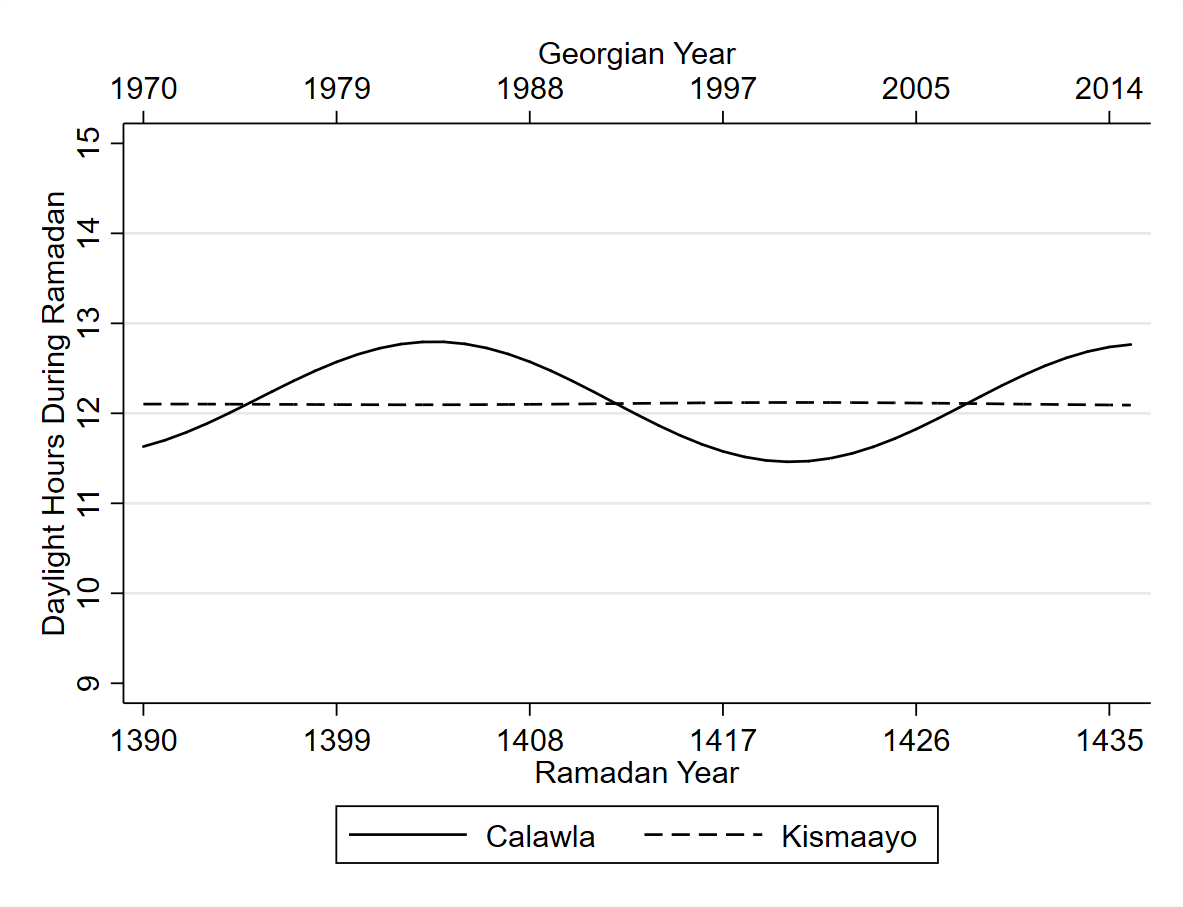}\caption{Somalia}
\end{subfigure}
\begin{subfigure}{0.475\textwidth}
\includegraphics[width=\textwidth]{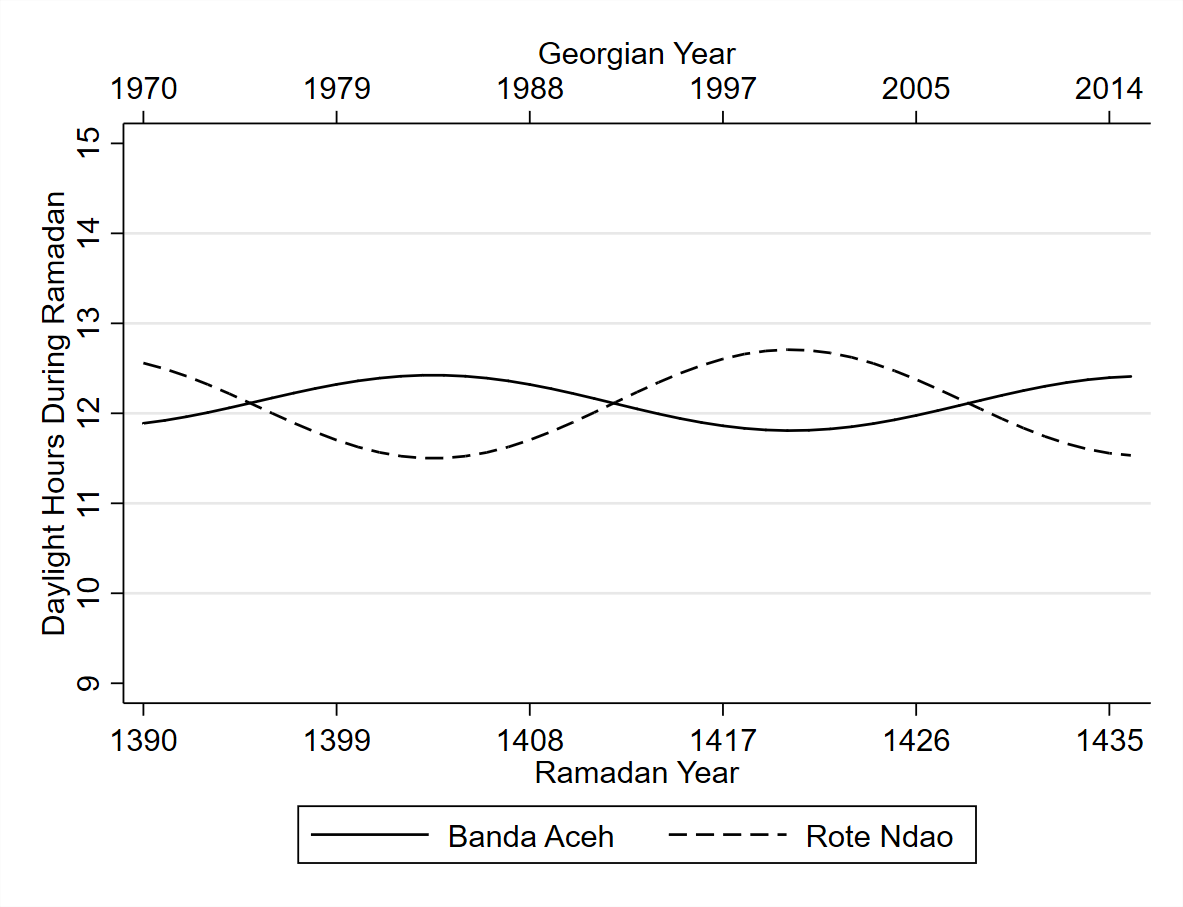}\caption{Indonesia}
\end{subfigure}
\end{figure}

Figure \ref{fig:cycle} shows the Ramadan daylight hours in the northernmost (solid lines) and southernmost (dashed lines) regions of Iraq, Pakistan, Somalia and Indonesia throughout our sample period.
\[ \text{Figure \ref{fig:cycle} around here}  \]
The Ramadan daylight hours oscillate around approximately 12 hours. The amplitude depends on the distance to the equator. The Ramadan daylight hours have larger amplitudes in districts farther away from the equator. Zakho is the northernmost district of Iraq and Al Faw the southernmost. The Ramadan daylight hours are longer in Zakho than in Al Faw when Ramadan is in the summer season. Conversely, Ramadan fasting is shorter in Zakho than in Al Faw when Ramadan is in the winter season. Ramadan daylight hours are similar for Iraq and Pakistan, in particular in their northernmost districts, which have similar latitudes. Ramadan daylight hours have smaller amplitudes in Somalia, which is much closer to the equator. They are almost time constant in the southernmost district of Somalia, Kismaayo, which is located almost on the equator. Ramadan daylight hours in the Indonesian district Banda Aceh and Rote Ndao oscillate in opposite directions. The reason is that Banda Aceh is in the northern hemisphere and Rote Ndao in the southern.

Comparing all 35 predominately Muslim countries and territories, we find that the maximum within-country difference in Ramadan daylight hours ranges from 1.2 hours in Indonesia to 6.3 hours in Uzbekistan, which is the northernmost of these countries. The maximum variation within country and Ramadan year varies from a few minutes in some small countries to around an hour (40--80 minutes) in Afghanistan, Algeria, Indonesia, Iran, Libya, Mali, Mauritania, Morocco, Oman, Pakistan, Saudi Arabia, and Somalia (see Table B.3 in Online Appendix B).

Ramadan daylight hours are exogenous after controlling for the latitude and the seasonality of Ramadan. We can account for these two factors with district and Ramadan year fixed effects. The district fixed effects capture factors that are time constant within districts, such as geography, culture and historical heritage. The Ramadan year fixed effects capture the Gregorian month during which Ramadan takes place in a given Ramadan year, as well as yearly varying factors that affect all countries equally. Possible candidates for such factors might be the global business cycle and tectonic shifts in geopolitics, such as the end of the Cold War, the September 11 attacks or the U.S.-led invasions in Afghanistan and Iraq. However, even such global factors are likely to affect different countries differently.

In addition, the literature has proposed a large number of potential determinants of terrorism that vary across countries and over time. For example, \cite{gass11} perform an extreme bound analysis in which they consider 65 correlates of terrorism at the country-year level that had been proposed by previous studies. Moreover, \cite{mei13} and \cite{end16} show that some of these determinants have heterogeneous effects across countries and time periods. For these reasons, we use interacted country-Ramadan year fixed effects instead of Ramadan year fixed effects. These interacted fixed effects allow us, among others, to implicitly control for the many previously proposed determinants of terrorism and to account for their potentially heterogeneous effects. Furthermore, they also control for country-level GDP and average happiness, which by themselves are affected by Ramadan fasting as shown by \cite{camp15}.\footnote{The findings by \cite{camp15} further imply that Ramadan fasting may affect terrorism differently in regions with longer Ramadan daylight hours because of larger economic losses in these regions. In an important robustness test, we thus control for economic activity as proxied by nighttime lights (see Section \ref{sec:results_main} and Table C.5 in Online Appendix C). We find no evidence that Ramadan fasting affects terrorism differently across regions with different Ramadan daylight hours because of differences in the economic effects of Ramadan fasting.}

Our main specification is
\begin{equation} \label{eq:main}
Terror_{ict}= \alpha_{i} + \beta_{ct} + \sum_{G=1}^{3} \gamma_{G} \left(RDH_{ict} \times 1_c\{c\in G\}\right) + \epsilon_{ict},
\end{equation}
where $Terror_{ict}$ is a terror outcome in district $i$ of country $c$ in Ramadan year $t$, e.g., the occurrence of at least one terrorist event. $\alpha_{i}$ and $\beta_{ct}$ represent the region and country-Ramadan year fixed effects mentioned above. $RDH_{ict}$ measures the average Ramadan daylight hours in region $i$ of country $c$ in Ramadan year $t$. $1_c\{c\in G\}$ represents indicator variables for the three mutually exclusive country groups $G=1,2,3$, which we stratify by the Muslim population share using thresholds of 25\% and 75\%. The indicator function $1_c\{c\in G\}$ is equal to one if country $c$ is part of group $G$, and zero otherwise. The parameters of interest are $\gamma_1$, $\gamma_2$ and $\gamma_3$. They measure the average effect of Ramadan daylight hours on terror outcomes within these three country groups. It is worth highlighting that we exploit only variation within countries and Ramadan years to identify these parameters. That is, in Figure \ref{fig:cycle} we exploit only the vertical variation in the Ramadan fasting duration within country-years, but not the horizontal variation across Ramadan years or any cross-country variation. This choice makes it a conservative specification.

The error term $\epsilon_{ict}$ absorbs unexplained variation of the terror outcome. We cluster the standard errors of the estimated coefficients at the country-level.

%%%%%%%%%%%%%%%%%%%%%%%%%%%%%%%%%%%%%%%%%%%%%%%%%%%%%%%%%%%%%%%%
%\newpage
\section{Results}\label{sec:results}

\subsection{Main results}\label{sec:results_main}

Table \ref{tab:main} reports the effects of the daily fasting duration during Ramadan, measured by Ramadan daylight hours, on the extensive margin of terrorist events and terror deaths in panels A and B, respectively. 
\[ \text{Table \ref{tab:main} around here}  \]
Column (1) looks at average effects and accounts separately for country and Ramadan year fixed effects. We find negative relations between Ramadan daylight hours and the two measures for the occurrence of terrorist events. These relations are not statistically significant at conventional levels. Column (2) includes district fixed effects as well as interacted country-Ramadan year fixed effects to account for constant region-specific characteristics and country-level changes of all sorts over time. The coefficient estimates and the standard errors remain similar. Hence, averaged across the globe, Ramadan daylight hours have no causal effect on the occurrence of terrorism. 

Column (3) introduces an interaction term between Ramadan daylight hours and the Muslim population share. The coefficient estimates on this interaction term are negative, suggesting a negative relation between the Muslim population share and the effect of Ramadan daylight hours on the occurrence of terrorist events. This coefficient, however, is not statistically significant at conventional levels. Moreover, the model in column (3) would be misspecified if the relation between the Muslim population share and the effect of Ramadan daylight hours were non-linear. 

We therefore turn to our preferred specification in column (4). There, we use three interaction terms between Ramadan daylight hours and indicator variables for whether the Muslim population share is below 25\%, between 25\% and 75\%, or above 75\%, respectively. The point estimates indeed hint at a non-linear relation. The effect of Ramadan daylight hours on the occurrence of terrorist events is negative, but small and not statistically significant in predominantly non-Muslim countries; positive, but imprecisely estimated in religiously divided countries;\footnote{Here and elsewhere, we refrain from interpreting the results for the (relatively few) religiously divided countries, because these positive effect is entirely driven by Nigeria (see Table C.11 in Online Appendix C).} and negative, sizable and statistically significant at the 5\%-level in predominantly Muslim countries. The point estimates suggest that an additional hour of daily fasting during Ramadan lowers the probability of a terrorist event in a predominantly Muslim country within 12 months after the beginning of Ramadan by 2.7 percentage points, and the probability of a fatal terrorist event by 2.4 percentage points. 

These results provide strong confirmation for our main hypothesis (H1). Moreover, they imply that the geography of terrorism depends on the timing of Ramadan within the solar cycle. Suppose Ramadan daylight hours differ by 30 minutes between a northern and a southern district of a predominantly Muslim country when Ramadan is in summer (around late June) or winter (around late December), which is approximately true for Iraq and Pakistan (see Figure \ref{fig:cycle}). Further suppose that these two districts have the same time-averaged propensity to terrorism, such that the probability of terrorist events is the same in these two districts if Ramadan is around late March or late September. Then, the probability of a terrorist event would be 1.35 percentage points lower in the northern (southern) than in the southern (northern) district if Ramadan were in summer (winter).\footnote{In comparison, the probability of a terrorist event occurring in an average year and district is 4.1\% across all predominantly Muslim countries, 14.2\% in Iraq, and 38.2\% in Pakistan.}

Online Appendix C present many robustness tests. We briefly discuss them, thereby focusing on our main results reported in column (4) of Table \ref{tab:main}. 
We start by studying the effect of longer Ramadan fasting on the intensity (rather than the extensive margin) of terrorism across districts. Table C.1 presents the results for the log-modulus transformation of the numbers of terrorist events and terror deaths per district (multiplied by 100). The pattern is remarkably similar as the one seen in Table \ref{tab:main}. Again, the effect of Ramadan daylight hours is negative, sizable and statistically significant in predominantly Muslim countries. We can interpret the coefficient estimates as approximations of semi-elasticities.\footnote{The coefficient estimates could be interpreted as semi-elasticities if we did not add one before taking the logarithm of terrorist events or terror deaths. In that case, however, we would lose most observations.} They suggest that an additional hour of daily fasting during Ramadan lowers terrorist events and terror deaths by approximately 3.3\% and 4.6\%, respectively.

Table C.2 replaces the continuous variable for Ramadan daylight hours with a set of indicator variables for different durations (and estimates separate models for each of the three country groups). Accordingly, the specifications are fully non-parametric, and there are no concerns about the limited support of the outcome variables. The reference category is 11.5--12.5 Ramadan daylight hours, capturing fasting durations close to the mean. We find that our main results are driven by particularly long Ramadan fasting. That is, there is a strong decrease in terrorist events in predominantly Muslim countries when Ramadan fasting is particularly long, but no corresponding increase when Ramadan fasting is particularly short. %The positive (and often imprecisely estimated) effects of Ramadan daylight hours on terrorist events in religiously divided countries are also driven by particularly long Ramadan fasting.

Table C.3 uses alternative thresholds of the Muslim population share to stratify the countries and territories into three groups. We find that the estimated effects for predominantly Muslim and predominantly non-Muslim countries remain very similar, while the imprecisely estimated point estimates for the countries with an intermediate Muslim population share are quite sensitive to the chosen thresholds.%\footnote{This pattern is not surprising as most countries have a Muslim population share below 5\% or above 95\% (see Figure A.2 in Online Appendix A). Therefore, the composition of the country groups with high or low Muslim population shares does not strongly depend on the exact thresholds, while the composition of the country group with an intermediate Muslim population share is fairly sensitive to the exact thresholds.}

Table C.4 focuses on the population share of the largest Islamic sect rather than on the population share of all Muslims aggregated across sects. It is based on information by the \href{http://www.pewforum.org/files/2009/10/Shiarange.pdf}{Pew Research Center} about the country-level population shares of Sunni and Shia Muslims. The estimated effects for countries where the larger of these two sects has a population share above 75\% are very similar to the effects for countries with a total Muslim population share above 75\% (reported in Table \ref{tab:main}).

Table C.5 adds time-varying district-level controls for economic activity and population size. Our measure of a district's economic activity is based on satellite data on the intensity of nighttime lights, provided by the National Oceanic and Atmospheric Administration (NOAA).\footnote{\cite{hend2012} and \cite{hr14} document a high correlation between changes in nighttime light intensity and GDP at the level of countries and subnational administrative regions, respectively. In addition, \cite{brue18} show that nighttime lights are correlated with broad measures of local social development as well.} The data come on a scale from 0 to 63 and in pixels of less than one square kilometer, which allows computing average nighttime light intensity within district boundaries. Our measure of a district's total population is based on population data from the Center for International Earth Science Information Network (CIESIN). Using these data, especially the nighttime lights, which are only available for the (Gregorian) years 1992--2013, reduces the sample size considerably. We find that controlling for these measures of local economic activity and local population does not change the sizable negative effects of Ramadan daylight hours on the probability of terrorists events occuring in predominantly Muslim countries. These results suggest that Ramadan fasting does not affect terrorism differently across regions with different Ramadan daylight hours because of differences in the economic effects of Ramadan fasting.

Table C.6 uses administrative regions at the first subnational level, typically provinces or states, as cross-sectional units. An average province or state consists of 13 districts. The coefficient estimates again suggest a large negative effect of Ramadan daylight hours on terrorists events in predominantly Muslim countries, but the standard errors become considerably larger.

Table C.7 disaggregate terrorist events by different sample periods: The time prior to the end of the Cold War (Ramadan years 1390--1411); the time between the end of the Cold War and the September 11 terrorist attacks (Ramadan years 1412--1422); and the years after the September 11 attacks (Ramadan years 1422-1436).\footnote{The first and the second Iraq wars started in close temporal proximity to the end of the Cold War and the September 11 attacks, respectively. Hence, results look very similar when splitting the sample into three time periods based on the beginnings of these wars, which have been very important to the Muslim world.} We find that the negative effect of longer Ramadan fasting on terrorist events in predominantly Muslim countries is driven by the time periods since the end of the Cold War. This finding is not surprising, as terrorist events were rare in these countries in earlier years (see Figure B.2 in Online Appendix B).
%OLD: Focusing on the first of these time periods, column (1) shows that longer Ramadan daylight hours had no effect on terrorist events, not even in predominantly Muslim countries. This finding is not surprising, given that terrorist events were rare in these countries prior to the end of the Cold War (see Figure \ref{fig:dev_Mshare}). Columns (2) and (3) show that longer Ramadan daylight hours reduced the probability of terrorist events in predominantly Muslim countries after the end of the Cold War, and that this effect became slightly larger after the September 11 terrorist attacks. These latter columns also shed light on the effects of longer Ramadan daylight hours on the two other country groups: First, the small negative average effect on the probability of terrorist events in predominantly non-Muslim countries is driven by the years since the September 2011 attacks. Second, the sizable, but imprecisely estimated positive average effect on the probability of terrorist events in religiously divided countries is driven by the entire time period since the end of the Cold War.

Table C.8 reports results after dropping Ramadan year 1412, which ended on February 22, 1993, and Ramadan year 1413, which started on February 23, 1993. The coefficient estimates and standard errors are almost identical as in Table \ref{tab:main}, column (4). Hence, our results are not an artefact of the missing data for the Gregorian year 1993 in the GTD.

Table C.9 reports results when including the 22,168 ambiguous events that START classifies to be acts of terrorism with high probability, but not certainty. Our results remain similar. If anything, the negative effects of longer Ramadan daylight hours on terrorist events and terror deaths in predominantly Muslim countries become even more pronounced.

Finally, Table C.10 reports separate results for predominantly Muslim countries in Africa and Asia. It further splits the Asian Muslim countries into Arab ones and the rest (proxied by membership in the Arab League). The negative effects of longer Ramadan daylight hours are particularly strong in Africa and the Arab world, and relatively weak elsewhere in Asia. Relatedly, Table C.11 reports results when re-estimating the main specification after sequentially dropping the five countries with most terror deaths during the sample period, i.e., Afghanistan, India, Iraq, Nigeria and Pakistan, as well as Israel and Palestine, and the United States, which are other prominent countries/territories in public debates on terrorism.\footnote{According to our data, the Muslim population share is below 25\% in India, Israel and the United States; between 25\% and 75\% in Nigeria; and above 75\% in Afghanistan, Iraq, Pakistan and Palestine.} The estimated effects for predominantly Muslim and predominantly non-Muslim countries remain similar in all cases. The imprecisely estimated positive effect for countries with an intermediate Muslim population share (see Table \ref{tab:main}), however, disappears when dropping Nigeria, which has the most terrorist events and terror deaths of all religiously divided countries.

The general pattern emerging is that Ramadan daylight hours have a robust negative effect on terrorist events in predominantly Muslim countries. By contrast, Ramadan daylight hours have no robust effect on terrorism in other countries.

%%%%%%%%%%%%%%%%%%%%%%%%%%%%%%%%%%%%%%%%%%%%%%%%%%%%%%%%%%%%%%%%
\subsection{Heterogeneous effects}\label{sec:results_hetero}

We next turn to our secondary hypothesis (H2), which predicts differential effects across attack types and targets. In columns (1) and (2) of Table \ref{tab:hetero}, we distinguish between suicide attacks and other/non-suicide attack types. We see that longer Ramadan daylight hours have a strong and statistically significant negative effect on non-suicide attacks in predominantly Muslim countries, but not on suicide attacks.
\[ \text{Table \ref{tab:hetero}  around here}  \]

In columns (3) and (4), we report the effects of Ramadan daylight hours on terrorist attacks against unarmed civilians and armed targets, respectively. We see that longer Ramadan daylight hours tend to reduce terrorist attacks on both types of targets in predominantly Muslim countries. The effect on the probability of a terrorist attack against civilian targets, however, is not statistically significant (and the effect on the probability of a fatal terrorist attack against civilian targets is only statistically significant at the 10\%-level). Comparing the point estimates and the sample means (reported in Table \ref{tab:des_panel}) further suggests that the relative change in terrorist attacks against civilian targets is smaller than the relative change in terrorist attacks against armed targets. 

Taken together, the results in columns (1)--(4) suggest that longer daily fasting during Ramadan fasting has a large negative effects on operationally difficult terrorist attacks in predominantly Muslim countries, but comparatively smaller and less robust negative effects on operationally easier attacks, such as suicide attacks and attacks against civilians. These findings confirm our second hypothesis (H2).

In columns (5) and (6), we look at differential effects of Ramadan fasting hours on terrorist attacks committed by Muslim and non-Muslim perpetrator groups. The results from this exercise need to be interpreted with caution, as we could only classify the perpetrators of 51\% of the terrorist events. This split nevertheless offers an interesting falsification test: If Ramadan daylight hours matter for terrorism only because they affect public support for Muslim perpetrator groups, then we should not find any effect on terrorist attacks committed by non-Muslim groups. Our estimates confirm that longer Ramadan daylight hours have no effect on terrorist attacks by non-Muslim perpetrators in any country group, but lead to significantly fewer terrorist attacks by Muslim perpetrators in predominantly Muslim countries.

%%%%%%%%%%%%%%%%%%%%%%%%%%%%%%%%%%%%%%%%%%%%%%%%%%%%%%%%%%%%%%%%
\subsection{Effects on public support for terrorism}\label{sec:results_ps}

A key element of our theoretical argument has been the assumption that more intense religious experiences reduce public support for terrorism in predominantly Muslim countries. We now test this prediction using survey data from the Pew Research Center. In particular, we use responses from Muslims who are asked whether they feel that violence is ``justified in order to defend Islam from its enemies.'' They can respond that such violence is ``never,'' ``rarely,'' ``sometimes''  or ``often''  justified, or they can refuse to answer. We take these responses as a proxy for the public support for terrorism. The Pew Research Center has included this question in surveys in 13 predominantly Muslim countries and territories since 2002. 57\% of the respondents feel that violence is never justified to defend Islam from its enemies. The corresponding shares who feel that it is rarely, sometimes or often justified are 13\%, 11\% and 9\%, respectively. 

To estimate the effect of longer Ramadan daylight hours on these shares, we employ the same empirical strategy as before, but the unit of observation is now a Muslim respondent from a particular province surveyed in a particular Ramadan year. Table \ref{tab:support} reports the results. 
\[ \text{Table \ref{tab:support} around here} \]
Column (1) suggests a large positive effect of longer Ramadan daylight hours on the share of people who feel that violence is never justified to defend Islam from its enemies, but the coefficient estimate is imprecisely estimated.
In contrast, columns (3) and (4) show that longer Ramadan daylight hours lead to a large and statistically significant reduction in the shares of individuals who feel that such violence is often or sometimes justified. This finding supports our assumption that more intense religious experiences reduce public support for terrorism in predominantly Muslim countries.\footnote{Online Appendix D presents results on the effects of Ramadan daylight hours on religious practices based on the survey data from the Pew Research Center. In particular, Tables D.1 and D.2 show that the longer Ramadan daylight hours tend to increase the share of respondents who fast all days during Ramadan as well as the share who pray five times per day rather than only weekly.}

%%%%%%%%%%%%%%%%%%%%%%%%%%%%%%%%%%%%%%%%%%%%%%%%%%%%%%%%%%%%%%%%
\subsection{Short-term physiological and psychological effects of fasting}\label{sec:dis_short}

The results presented in Sections \ref{sec:results_main}--\ref{sec:results_ps} provide strong empirical support for our theoretical argument. In this section, we nevertheless evaluate some alternative channels through which longer Ramadan daylight hours may impact on terrorism. Medical studies document a plethora of physiological and psychological effects of Ramadan fasting, including body weight loss, sleep deprivation, tiredness, lassitude and irritability \citep[e.g.,][]{lei03}. Longer Ramadan fasting could lower terrorism due to such short-term physiological and psychological effects on  (potential) terrorists. Alternatively, it could also impact on terrorism by affecting the behavior of security forces. We know that terror events tend to be neither more, nor less common in Ramadan than in other months (see Figures B.1--B.3 in Online Appendix B). But to properly test whether physiological and psychological effects are a plausible channel through which longer Ramadan fasting hours could decrease terrorist attacks, we make use of the fact that physiological and psychological consequences of fasting should be temporary. Hence, if we find that longer Ramadan fasting hours have an effect on terrorism beyond the month of Ramadan and the next few weeks or months, then that suggests that physiological and psychological effects are unlikely to be the main channel.

We construct a new panel dataset with the same cross-sectional units but quarterly frequency instead of annual frequency. The first quarter of each Ramadan year starts with the month of Ramadan and also includes the two subsequent Islamic months. The second quarter contains the next three Islamic months. The third quarter and the fourth are constructed analogously. Tables E.1 and Table E.2 in Online Appendix E present descriptive statistics and our estimates for the quarterly panel dataset. As expected, we find that longer Ramadan daylight hours lead to a considerable decrease in terrorism during the first quarter of a Ramadan year in predominantly Muslim countries. Moreover, this effect is fairly persistent and does not become much weaker in subsequent quarters of the same Ramadan year.\footnote{This finding is based on specification that include triple interaction terms between (i) Ramadan daylight hours, (ii) the indicator variable for countries with a Muslim population share above 75\%, and (iii) indicator variables for the second, third and fourth quarters of a Ramadan year. The coefficient estimates of these triple interaction terms are small in absolute value and typically not statistically significant (see columns (2) and (5) of Table D.2 in Online Appendix D). These results are robust to the addition of country-Ramadan year quarter fixed effects (see columns (3) and (6) of Table D.2 in Online Appendix D).} Hence, short-term physiological and psychological effects of fasting are very unlikely to be the main drivers of our results.

%\footnote{As seen in Section \ref{sec:results}, longer Ramadan daylight hours do not reduce (and may even increase) terrorist events in countries with a Muslim population share between 25\% and 75\%. The question arises as to why the proposed channel is not at work in these countries. A possible reason is related to the effect of Ramadan fasting duration on the views of Muslims on non-Muslims. Table E.1 in Online Appendix E presents results based on survey questions on the views on Christians and Jews in six predominately Muslim countries. It shows that longer Ramadan daylight hours increase the share of Muslim respondents with unfavorable views on Christians and Jews. Therefore, some Muslims seem to feel more at unease with both terrorism and non-Muslims after intense Ramadan fasting. It is conceivable that these countervailing effects are one of the reasons why longer Ramadan fasting does not lead to a reduction in terrorist events in religiously divided societies.}

%%%%%%%%%%%%%%%%%%%%%%%%%%%%%%%%%%%%%%%%%%%%%%%%%%%%%%%%%%%%%%%%
%\newpage
\section{Concluding remarks}\label{sec:con}

The long history of terrorist groups with a religious background and the recent surge of terrorist attacks by Islamist perpetrators in predominantly Muslim countries has motivated us to study the effect of religion and intense religious experiences on terrorism. We have focused on one of the five pillars of Islam: Daily fasting from dawn to sunset during the month of Ramadan. We have developed a theoretical argument based on the extensive literature documenting the importance of public support for the success of terrorist groups, and our assumption that more intensive Ramadan fasting would reduce the public support for terrorism. Our argument predicts that longer Ramadan fasting hours -- by reducing public support for terrorism and the operational capability of terrorist groups -- result in less terrorist attacks in Muslim countries, and that this effect is stronger for operationally more difficult attacks.

Building on \cite{camp15}, we have exploited the facts that the daily fast in Ramadan lasts from dawn to sunset and that the lunar Islamic calendar is not synchronized with the solar cycle. We have identified causal effects by focusing on differences in Ramadan fasting hours across districts within country-years. We have found evidence in support of our assumption that longer Ramadan fasting hours reduce public support for terrorism. We have presented strong empirical support for our main hypothesis that longer Ramadan fasting hours reduce terrorism in predominantly Muslim countries. In addition, confirming our second hypothesis, we have found that the negative effect of longer Ramadan fasting hours are stronger for operationally difficult terrorist attacks than for suicide attacks or attacks against unarmed civilians.

Hence, while Muslim terrorist groups may (mis-)use religion to (self-)legitimize their attacks and to try to gain public support, it would be wrong to think that intense religious experiences are necessarily a breeding ground for terrorism. Quite to the contrary, our results show that intense religious experiences can reduce public support for terrorism and, thereby, terrorist attacks.

%One could interpret our findings as support for the promotion of peaceful interpretations of the Quran as counter-narratives to the interpretation and propaganda of Islamist groups. It is, however, questionable whether counter-narratives promoted by foreigners or a domestic elite can have similar effects as higher religiosity due to more intense Ramadan fasting. This question is related to the general concern -- forcefully articulated by \cite{sen07} -- that counter-narratives may play into the hands of Islamist groups by focusing on religious identity rather than the plurality of identities each individual has.

\onehalfspacing
%%%%%%%%%%%%%%%%%%%%%%%%%%%%%%%%%%%%%%%%%%%%%%%%%%%%%%%%%%%%%%%%
%\clearpage
\bibliographystyle{ecca} 
\bibliography{Bibliothek}

\begin{thebibliography}{28}
\providecommand{\natexlab}[1]{#1}

\bibitem[{Atran(2003)}]{atr03}
\textsc{Atran, S.} (2003). {Genesis of Suicide Terrorism}. \textit{Science},
  \textbf{299}~(5612), 1534--1539.

\bibitem[{Berman(2011)}]{ber11}
\textsc{Berman, E.} (2011). \textit{{Radical, Religious, and Violent: The New
  Economics of Terrorism}}. MIT Press.

\bibitem[{Berman and Laitin(2008)}]{ber08}
\textsc{---} and \textsc{Laitin, D.} (2008). {Religion, Terrorism and Public
  Goods: Testing the Club Model}. \textit{Journal of Public Economics},
  \textbf{92}~(10-11), 1942--1967.

\bibitem[{Bruederle and Hodler(2018)}]{brue18}
\textsc{Bruederle, A.} and \textsc{Hodler, R.} (2018). {Nighttime Lights as a
  Proxy for Human Development at the Local Level}. \textit{PLOS One},
  \textbf{13}~(9), e0202231.

\bibitem[{Bueno~de Mesquita and Dickson(2007)}]{buen07}
\textsc{Bueno~de Mesquita, E.} and \textsc{Dickson, E.~S.} (2007). {The
  Propaganda of the Deed: Terrorism, Counterterrorism, and Mobilization}.
  \textit{American Journal of Political Science}, \textbf{51}~(2), 364--381.

\bibitem[{Campante and Yanagizawa-Drott(2015)}]{camp15}
\textsc{Campante, F.} and \textsc{Yanagizawa-Drott, D.} (2015). {Does Religion
  Affect Economic Growth and Happiness? Evidence from Ramadan}.
  \textit{Quarterly Journal of Economics}, \textbf{130}~(1), 615--658.

\bibitem[{Clingingsmith \textit{et~al.}(2009)Clingingsmith, Khwaja and
  Kremer}]{cli09}
\textsc{Clingingsmith, D.}, \textsc{Khwaja, A.~I.} and \textsc{Kremer, M.}
  (2009). {Estimating the Impact of the Hajj: Religion and Tolerance in Islam's
  Global Gathering}. \textit{Quarterly Journal of Economics}, \textbf{124}~(3),
  1133--1170.

\bibitem[{Comerford and Bryson(2017)}]{com17}
\textsc{Comerford, M.} and \textsc{Bryson, R.} (2017). \textit{{Struggle Over
  Scripture Charting the Rift Between Islamist Extremism and Mainstream
  Islam}}. Tony Blair Institute for Global Change.

\bibitem[{Enders \textit{et~al.}(2016)Enders, Hoover and Sandler}]{end16}
\textsc{Enders, W.}, \textsc{Hoover, G.~A.} and \textsc{Sandler, T.} (2016).
  {The Changing Nonlinear Relationship between Income and Terrorism}.
  \textit{Journal of Conflict Resolution}, \textbf{60}~(2), 195--225.

\bibitem[{Gassebner and Luechinger(2011)}]{gass11}
\textsc{Gassebner, M.} and \textsc{Luechinger, S.} (2011). {Lock, Stock, and
  Barrel: A Comprehensive Assessment of the Determinants of Terror}.
  \textit{Public Choice}, \textbf{149}~(3-4), 235.

\bibitem[{Haruvy \textit{et~al.}(2018)Haruvy, Ioannou and Golshirazi}]{har18}
\textsc{Haruvy, E.~E.}, \textsc{Ioannou, C.~A.} and \textsc{Golshirazi, F.}
  (2018). {The Religious Observance of Ramadan and Prosocial Behavior}.
  \textit{Economic Inquiry}, \textbf{56}~(1), 226--237.

\bibitem[{Henderson \textit{et~al.}(2012)Henderson, Storeygard and
  Weil}]{hend2012}
\textsc{Henderson, J.~V.}, \textsc{Storeygard, A.} and \textsc{Weil, D.~N.}
  (2012). {Measuring Economic Growth from Outer Space}. \textit{American
  Economic Review}, \textbf{102}~(2), 994--1028.

\bibitem[{Hodler and Raschky(2014)}]{hr14}
\textsc{Hodler, R.} and \textsc{Raschky, P.~A.} (2014). {Regional Favoritism}.
  \textit{Quarterly Journal of Economics}, \textbf{129}~(2), 995--1033.

\bibitem[{Iannaccone(1992)}]{ian92}
\textsc{Iannaccone, L.} (1992). {Sacrifice and Stigma: Reducing Free-riding in
  Cults, Communes, and Other Collectives}. \textit{Journal of Political
  Economy}, \textbf{100}~(2), 271--291.

\bibitem[{Jaeger \textit{et~al.}(2015)Jaeger, Klor, Miaari and
  Paserman}]{dav15}
\textsc{Jaeger, D.~A.}, \textsc{Klor, E.}, \textsc{Miaari, S.} and
  \textsc{Paserman, M.~D.} (2015). {Can Militants Use Violence to Win Public
  Support? Evidence from the Second Intifada}. \textit{Journal of Conflict
  Resolution}, \textbf{58}~(3), 528--549.

\bibitem[{Kis-Katos \textit{et~al.}(2014)Kis-Katos, Liebert and
  Schulze}]{kis14}
\textsc{Kis-Katos, K.}, \textsc{Liebert, H.} and \textsc{Schulze, G.~G.}
  (2014). {On the Heterogeneity of Terror}. \textit{European Economic Review},
  \textbf{68}, 116--136.

\bibitem[{Krueger and Male\v{c}kov\'{a}(2009)}]{kru09}
\textsc{Krueger, A.} and \textsc{Male\v{c}kov\'{a}, J.} (2009). {Attitudes and
  Action: Public Opinion and the Occurrence of International Terrorism}.
  \textit{Science}, \textbf{325}~(5947), 1534--1536.

\bibitem[{Leiper and Molla(2003)}]{lei03}
\textsc{Leiper, J.~B.} and \textsc{Molla, A.} (2003). {Effects on Health of
  Fluid Restriction during Fasting in Ramadan}. \textit{European Journal of
  Clinical Nutrition}, \textbf{57}~(S2), S30.

\bibitem[{Male\v{c}kov\'{a} and Stani\v{s}i\'{c}(2011)}]{mal11}
\textsc{Male\v{c}kov\'{a}, J.} and \textsc{Stani\v{s}i\'{c}, D.} (2011).
  {Public Opinion and Terrorist Acts}. \textit{European Journal of Political
  Economy}, \textbf{27}~(1), S107--S121.

\bibitem[{Male\v{c}kov\'{a} and Stani\v{s}i\'{c}(2013)}]{mal13}
\textsc{---} and \textsc{Stani\v{s}i\'{c}, D.} (2013). {Does Higher Education
  Decrease Support for Terrorism?} \textit{Peace Economics, Peace Science and
  Public Policy}, \textbf{19}~(3), 343--358.

\bibitem[{Meierrieks and Gries(2013)}]{mei13}
\textsc{Meierrieks, D.} and \textsc{Gries, T.} (2013). {Causality Between
  Terrorism and Economic Growth}. \textit{Journal of Peace Research},
  \textbf{50}~(1), 91--104.

\bibitem[{Polo and Gleditsch(2016)}]{polo16}
\textsc{Polo, S.~M.} and \textsc{Gleditsch, K.~S.} (2016). {Twisting Arms and
  Sending Messages: Terrorist Tactics in Civil War}. \textit{Journal of Peace
  Research}, \textbf{53}~(6), 815--829.

\bibitem[{Reese \textit{et~al.}(2017)Reese, Ruby and Pape}]{reese2017}
\textsc{Reese, M.~J.}, \textsc{Ruby, K.~G.} and \textsc{Pape, R.~A.} (2017).
  {Days of Action or Restraint? How the Islamic Calendar Impacts Violence}.
  \textit{American Political Science Review}, \textbf{11}~(3), 439--459.

\bibitem[{Siqueira and Sandler(2006)}]{siqu06}
\textsc{Siqueira, K.} and \textsc{Sandler, T.} (2006). {Terrorists versus the
  Government: Strategic Interaction, Support, and Sponsorship}. \textit{Journal
  of Conflict Resolution}, \textbf{50}~(6), 878--898.

\bibitem[{{START}(2017)}]{start_data}
\textsc{{START}} (2017). \textit{{Global Terrorism Database (GTD) [Datafile]}}.
  {National Consortium for the Study of Terrorism and Responses to Terrorism.
  Retrieved from https://www.start.umd.edu/gtd in summer 2017}.

\bibitem[{{START}(2018)}]{start_codebook}
\textsc{{START}} (2018). \textit{{Global Terrorism Database (GTD): Codebook}}.
  {National Consortium for the Study of Terrorism and Responses to Terrorism}.

\bibitem[{Tessler and Robbins(2007)}]{tes07}
\textsc{Tessler, M.} and \textsc{Robbins, M. D.~H.} (2007). {What Leads Some
  Ordinary Arab Men and Women to Approve of Terrorist Acts Against the United
  States?} \textit{Journal of Conflict Resolution}, \textbf{51}~(2), 305--328.

\bibitem[{Toft and Zhukov(2015)}]{toft15}
\textsc{Toft, M.~D.} and \textsc{Zhukov, Y.~M.} (2015). {Islamists and
  Nationalists: Rebel Motivation and Counterinsurgency in Russia's North
  Caucasus}. \textit{American Political Science Review}, \textbf{109}~(2),
  222--238.

\end{thebibliography}

%%%%%%%%%%%%%%%%%%%%%%%%%%%%%%%%%%%%%%%%%%%%%%%%%%%%%%%%%%%%%%%%
\clearpage
\section*{Figures and Tables}

\clearpage%%%%%%%%%%%%%%%%%%%%%%%%%%%%%%%%%%%%%%%%%%%%%%%%%%%%%%

\begin{table}[h!]
\centering
\caption{Ramadan fasting in Muslim countries} \label{tab:pew_fast}
\begin{tabular}{lcc}
\hline \hline
Country & Ramadan fasting: & Obs. \\
 & most/all days &  \\
\cline{2-3}     
 & (1)   & (2)   \\ 
\hline
Egypt       & 94\%  & 4,653 \\  
Indonesia   & 90\%  & 2,769 \\
Jordan      & 95\%  & 2,890 \\
Kuwait      & 99\%  & 471 \\
Mali        & 71\%  & 622 \\
Pakistan    & 78\%  & 9,244 \\
Palestine   & 93\%  & 3,540 \\
Senegal     & 85\%  & 742 \\
Tunisia     & 93\%  & 981 \\
Turkey      & 81\%  & 5,524 \\
\hline
Weighted average & 86\%  & 31,436 \\
\hline\hline
\end{tabular}
\parbox{0.53\textwidth}{\footnotesize \emph{Notes:} Results are based on Muslim respondents in surveys by the Pew Research Center from the years 2007--2013. Column (1) shows the share answering ``During most or all days of Ramadan'' or ``During all of Ramadan and other religious holidays'' as opposed to lower frequency when asked ``How often, if at all, do you fast?'' Column (2) provides the number of responses on which the share in column (1) is based.}
\end{table}

\newpage%%%%%%%%%%%%%%%%%%%%%%%%%%%%%%%%%%%%%%%%%%%%%%%%%%%%%%

\begin{table}[h!] \centering
\caption{Descriptive statistics of the district-Ramadan-year panel data, by country group} \label{tab:des_panel}
   %\begin{small}
   \begin{tabular}{lcccccc} \hline\hline
          & \multicolumn{6}{c}{Muslim population share:} \\
          & \multicolumn{2}{c}{$<$ 25\%} & \multicolumn{2}{c}{25--75\%} & \multicolumn{2}{c}{$>$75\%} \\
          & Mean  & Std.Dev. & Mean  & Std.Dev. & Mean  & Std Dev. \\
   \cline{2-7} & (1)   & (2)   & (3)   & (4)   & (5)   & (6) \\\hline
    Ramadan daylight hours & 12.24 & 1.81 & 12.18 & 0.93 & 12.26 & 1.37 \\
    Muslim pop.\ share (in \%)  & 2.28 & 3.90 & 48.45 & 8.75 & 96.25 & 5.24 \\ \hline
    \multicolumn{7}{c}{Terrorist events} \\\hline
    All events: &       &       &       &       &       &  \\
    \quad  Extensive margin & 1.32  & 11.40 & 1.77  & 13.18 & 4.09  & 19.80 \\
    \quad  Log-modulus  & 1.94  & 20.29 & 2.71  & 24.32 & 7.80  & 53.10 \\
    %\quad  Square-root & 1.41  & 14.15 & 1.97  & 17.11 & 5.42  & 31.79 \\
    \multicolumn{7}{l}{By perpetrator groups (extensive margins):} \\
    \quad  Muslim & 0.11  & 3.33  & 0.74  & 8.55  & 2.60  & 15.91 \\
    \quad  Non-Muslim & 0.79  & 8.86  & 0.25  & 4.96  & 0.30  & 5.46 \\
    %\quad  Islamist & 0.06  & 2.42  & 0.45  & 6.68  & 1.92  & 13.72 \\
    %\quad  Secular Muslim & 0.03  & 1.81  & 0.15  & 3.89  & 0.40  & 6.35 \\
    \multicolumn{7}{l}{By attack types (extensive margins):} \\
    \quad Suicide attacks & 0.01  & 1.22  & 0.11  & 3.28  & 0.62  & 7.83 \\
    \quad Other attacks &  1.31  & 11.38  &  1.74  &  13.08  & 4.00  & 19.59 \\
    %\quad  Assault & 0.46  & 6.75  & 0.78  & 8.78  & 1.91  & 13.68 \\
    %\quad  Bombing & 0.64  & 8.00  & 0.78  & 8.82  & 2.47  & 15.51 \\
    %\quad  Suicide attack & 0.01  & 1.22  & 0.11  & 3.28  & 0.62  & 7.83 \\
    %\quad  Assassination & 0.29  & 5.34  & 0.29  & 5.35  & 1.05  & 10.20 \\    
    \multicolumn{7}{l}{By targets (extensive margins):} \\
    \quad  Armed targets & 0.37  & 6.04  & 0.48  & 6.93  & 1.86  & 13.50 \\
    \quad  Unarmed civilians & 1.01  & 9.99  & 1.46  & 11.99 & 3.19  & 17.57 \\
    %\quad  Infrastructure & 0.27  & 5.19  & 0.25  & 4.97  & 0.76  & 8.71 \\
    \hline
    \multicolumn{7}{c}{Terror deaths} \\\hline
    All deaths: &       &       &       &       &       &  \\
    \quad  Extensive margin & 0.67  & 8.15  & 1.15  & 10.66 & 3.03  & 17.15 \\
    \quad  Log-modulus  & 1.52  & 24.33 & 4.06  & 53.53 & 10.04 & 82.97 \\
    %\quad  Square-root & 1.07  & 15.18 & 2.43  & 26.73 & 6.23  & 41.33 \\
    \multicolumn{7}{l}{By perpetrator groups (extensive margins):} \\
    \quad  Muslim & 0.06  & 2.51  & 0.57  & 7.54  & 2.00  & 14.00 \\
    \quad  Non-Muslim & 0.40  & 6.33  & 0.13  & 3.58  & 0.17  & 4.14 \\
    %\quad  Islamist & 0.04  & 1.87  & 0.37  & 6.11  & 1.55  & 12.37 \\
    %\quad  Secular Muslim & 0.01  & 1.18  & 0.08  & 2.83  & 0.25  & 5.04 \\
    \multicolumn{7}{l}{By attack types (extensive margins):} \\
    \quad Suicide attacks & 0.01  & 1.19  & 0.10  & 3.22  & 0.59  & 7.67 \\
    \quad Other attacks & 0.66  &  8.11  & 1.12  & 10.51  &  2.91  &  16.80 \\
    %\quad  Assault & 0.31  & 5.52  & 0.62  & 7.84  & 1.58  & 12.45 \\
    %\quad  Bombing & 0.20  & 4.43  & 0.41  & 6.41  & 1.64  & 12.69 \\
    %\quad  Suicide attack & 0.01  & 1.19  & 0.10  & 3.22  & 0.59  & 7.67 \\\hline
    %\quad  Assassination & 0.24  & 4.85  & 0.20  & 4.51  & 0.85  & 9.17 \\\hline
    \multicolumn{7}{l}{By targets (extensive margins):} \\
    \quad  Armed targets & 0.23  & 4.75  & 0.35  & 5.87  & 1.49  & 12.11 \\
    \quad  Civilians & 0.49  & 7.01  & 0.92  & 9.53  & 2.27  & 14.89 \\
    %\quad  Infrastructure & 0.06  & 2.52  & 0.08  & 2.88  & 0.33  & 5.76
    \hline
    Observations & \multicolumn{2}{c}{1,591,279} & \multicolumn{2}{c}{93,671} & \multicolumn{2}{c}{189,692} \\
    \hline \hline
	\end{tabular}
	%\end{small}
	\parbox{0.9\textwidth}{\footnotesize \emph{Notes:} Section \ref{sec:data} describes the data.}
\end{table} 

\newpage%%%%%%%%%%%%%%%%%%%%%%%%%%%%%%%%%%%%%%%%%%%%%%%%%%%%%%

\begin{table}[h!]  
\caption{Effects of longer Ramadan fasting on the occurrence of terrorist events and terror deaths} \label{tab:main}\centering
\begin{small}
\begin{tabular}{lcccc}\hline\hline
 & (1)   & (2)   & (3)   & (4) \\\hline
\multicolumn{5}{c}{A: Terrorist event (extensive margin)} \\\hline
Ramadan daylight hours          &      -0.077   &      -0.069   &       0.019   &               \\
            &     (0.073)   &     (0.084)   &     (0.100)   &               \\ 
\quad $\times$ Muslim pop.\ share     &               &               &      -1.686   &               \\
            &               &               &     (1.319)   &               \\
\quad $\times$ Muslims $<$25\%  &               &               &               &      -0.049   \\
            &               &               &               &     (0.075)   \\
\quad $\times$ Muslims 25-75\%  &               &               &               &       1.026   \\
            &               &               &               &     (1.208)   \\
\quad $\times$ Muslims $>$75\%  &               &               &               &      -2.704** \\
            &               &               &               &     (1.209)   \\
\hline
\multicolumn{5}{c}{B: Terror deaths (extensive margin)} \\\hline
Ramadan daylight hours          &      -0.040   &      -0.033   &       0.046   &               \\
            &     (0.044)   &     (0.049)   &     (0.068)   &               \\
\quad $\times$ Muslim pop.\ share     &               &               &      -1.504   &               \\
            &               &               &     (1.111)   &               \\
\quad $\times$ Muslims $<$25\%  &               &               &               &      -0.018   \\
            &               &               &               &     (0.040)   \\
\quad $\times$ Muslims 25-75\%  &               &               &               &       1.231   \\
            &               &               &               &     (1.389)   \\
\quad $\times$ Muslims $>$75\%  &               &               &               &      -2.359** \\
            &               &               &               &     (0.970)   \\
\hline
Country fixed effects (FE) & Yes & No & No & No \\
Ramadan year FE & Yes & No & No & No \\
District FE & No & Yes & Yes & Yes \\
Country-Ramadan year FE & No & Yes & Yes & Yes \\
Observations & 1,874,642 & 1,874,642 & 1,874,642 & 1,874,642 \\ \hline\hline
\end{tabular}
\end{small}
\parbox{\textwidth}{\footnotesize \emph{Notes:} Panel fixed effects regressions. Dependent variables are the extensive margins for terrorist events in panel A and terror deaths in panel B. Section \ref{sec:data} introduces the time unit ``Ramadan year'' and all the variables used. Standard errors are clustered at the country-level and reported in parentheses. ***/**/* indicate statistical significance at the 1\%/5\%/10\%-level.}
\end{table}    

\newpage

\begin{table}[h!]  
\caption{Heterogeneous effects of longer Ramadan fasting on terrorist events and terror deaths} \label{tab:hetero}\centering
\begin{tabular}{lcccccc}\hline\hline
 & \multicolumn{2}{c}{Attack type:} & \multicolumn{2}{c}{Target:} & \multicolumn{2}{c}{Perpetrator group:} \\
 & Suicide & Other & Unarmed & Armed & Non- & Muslim \\ 
 & attacks & attacks & civilians & targets & Muslim & \\
 & (1) & (2) & (3) & (4) & (5) & (6) \\\hline
\multicolumn{7}{c}{A: Terrorist event (extensive margin)} \\\hline
Ramadan daylight hours  & & \\
\quad $\times$ Muslims $<$25\%  &     0.002   &      -0.050  &      -0.044   &      -0.016   &      -0.030   &       0.032   \\
            &     (0.003)   &     (0.075)   &     (0.061)   &     (0.028)   &     (0.031)   &     (0.046)   \\
\quad $\times$ Muslims 25-75\%  &        0.474   &       0.955   &       1.116   &       0.727   &      -0.246   &       1.679   \\
            &       (0.489)   &     (1.119)   &     (1.068)   &     (0.771)   &     (0.308)   &     (1.589)   \\
\quad $\times$ Muslims $>$75\%  &       -0.220   &      -2.643** &      -1.496   &      -1.798***&       0.041   &      -2.306** \\
            &      (0.519)   &     (1.235)    &     (0.916)   &     (0.538)   &     (0.179)   &     (0.929)   \\
\hline
\multicolumn{7}{c}{B: Terror deaths (extensive margin)} \\\hline
Ramadan daylight hours  & & \\
\quad $\times$ Muslims $<$25\%  &       0.000   &      -0.016   &      -0.027   &      -0.000   &      -0.007   &       0.027   \\
            &     (0.002)   &     (0.041)    &     (0.031)   &     (0.015)   &     (0.019)   &     (0.034)   \\
\quad $\times$ Muslims 25-75\%  &        0.462   &       1.116   &       1.062   &       0.613   &      -0.102   &       1.640   \\
            &      (0.477)   &     (1.259)    &     (1.178)   &     (0.607)   &     (0.157)   &     (1.572)   \\
\quad $\times$ Muslims $>$75\%  &      -0.236   &      -2.340** &      -1.407*  &      -1.515***&      -0.007   &      -1.719** \\
            &       (0.524)   &     (0.961)    &     (0.782)   &     (0.453)   &     (0.109)   &     (0.769)   \\
\hline
District fixed effects (FE) & Yes & Yes & Yes & Yes & Yes & Yes \\
Country-Ramadan year FE & Yes & Yes & Yes & Yes & Yes & Yes \\
Observations & 1,874,642 & 1,874,642 & 1,874,642 & 1,874,642 & 1,874,642 & 1,874,642 \\ \hline\hline
\end{tabular}
\parbox{\textwidth}{\footnotesize \emph{Notes:} Panel regressions with district and country-Ramadan year fixed effects. Dependent variables are based on the extensive margins for terrorist events in panel A and terror deaths in panel B. They are restricted to suicide attacks in column (1), other/non-suicide attacks in column (2), terrorist events against unarmed civilians in column (3), terrorist events against armed targets in column (4), terrorist events by non-Muslim perpetrator groups in column (5), and terrorist events by Muslim perpetrator groups in column (6). Section \ref{sec:data} introduces the time unit ``Ramadan year'' and all the variables used. Standard errors are clustered at the country-level and reported in parentheses. ***/**/* indicate statistical significance at the 1\%/5\%/10\%-level.}
\end{table}

\newpage%%%%%%%%%%%%%%%%%%%%%%%%%%%%%%%%%%%%%%%%%%%%%%%%%%%%%%

\begin{table}[h!]
\centering
\caption{Effect of longer Ramadan fasting on public support for terrorism} \label{tab:support}
\begin{tabular}{lcccc}
\hline\hline
% & \multicolumn{4}{c}{Violence justified ``to defend Islam from its enemies?''} \\
 & \multicolumn{4}{c}{Violence justified against ``enemies'' of Islam?} \\
 & \, Never \,\, & \,\, Rarely \, & Sometimes & Often \\ 
 \cline{2-5}
 & (1) & (2)  & (3) & (4) \\ \hline
 
Ramadan daylight hours &    10.33         &     3.76         &    -5.41*** &    -8.98** \\
          &  (12.17)         &   (5.24)         &   (2.01)         &   (4.34)         \\ \hline
Province fixed effects (FE)	& Yes & Yes	& Yes & Yes \\
Country-Ramadan year FE	& Yes & Yes	& Yes & Yes \\
%Mean dep.\ var. & 9.15 & 11.22 & 13.50 & 57.40 \\ 
Observations &    52,452         &    52,452         &    52,452         &    52,452         \\ \hline\hline
\end{tabular}
\parbox{\textwidth}{\footnotesize \emph{Notes:} Panel regressions with province and country-Ramadan year fixed effects. The sample is based on Pew surveys conducted by the Pew Research Center in 13 predominantly Muslim countries and territories (Bangladesh, Egypt, Indonesia, Jordan, Kuwait, Mali, Morocco, Pakistan, Palestine, Senegal, Tunisia, Turkey, and Uzbekistan) in the years 2002--2015. 
%It was asked in multiple surveys in Egypt, Indonesia, Jordan, Pakistan, Palestine, Senegal, Tunisia and Turkey.
The dependent variables are equal to 100 if the Muslim respondent gave the answer indicated at the top of each column (never, rarely, sometimes, often) when asked whether they personally feel that ``suicide bombing and other forms of violence against civilian targets are justified in order to defend Islam from its enemies,'' and zero otherwise
Standard errors in parentheses are clustered at the level of provinces. ***/**/* indicate statistical significance at the 1\%/5\%/10\%-level.}
\end{table}

\clearpage
\setcounter{page}{1}
\setcounter{footnote}{0}
\setcounter{equation}{0}
\begin{center}
{\LARGE Online Appendix to ``Religion and Terrorism: \\ \medskip Evidence from Ramadan Fasting''} \bigskip \bigskip

{\large Roland Hodler, Paul A.\ Raschky and Anthony Strittmatter\footnote{Hodler: Department of Economics, University of St.Gallen; CEPR; CESifo; email: roland.hodler@unisg.ch. Raschky: Department of Economics, Monash University; email: paul.raschky@monash.edu. Strittmatter: Department of Economics, University of St.Gallen; email: anthony.strittmatter@unisg.ch.}} \bigskip
\end{center}

\subsection*{Sections:}
\begin{enumerate}
\item[A.] List of countries and territories, stratified by Muslim population shares
\item[B.] Additional descriptive statistics 
\item[C.] Robustness tests 
\item[D.] Effects of Ramadan daylight hours on religious practices 
\item[E.] Summary statistics and results for quarterly panel
%\item[E.] Ramadan fasting and views on Christians and Jews
\end{enumerate}

%%%%%%%%%%%%%%%%%%%%%%%%%%%%%%%%%%%%%%%%%%%%%%%%%%%%%%%%%%%%%%%%%%%%%%%%%%%%%%%%%%%%%
\clearpage 
\newpage
\section*{A. List of all countries and territories, stratified by Muslim population shares}

This appendix lists all countries and territories in our sample and stratifies them into three groups based on their Muslim population share (see  Section \ref{sec:data}).

\paragraph{Countries and territories with a Muslim population share below 25\%:} 
American Samoa, Angola, Antigua and Barbuda, Argentina, Armenia, Australia, Austria, Barbados, Belarus, Belgium, Belize, Benin, Bermuda, Bhutan, Bolivia, Botswana, Brazil, Bulgaria, Burundi, Cambodia, Cameroon, Canada, Cape Verde, Cayman Islands, Central African Republic, Chile, China, Colombia, Costa Rica, Croatia, Cuba, Cyprus, Czech Republic, Dem.\ Rep.\ of Congo, Denmark, Dominica, Dominican Republic, East Timor, Ecuador, El Salvador, Equatorial Guinea, Estonia, Faroe Islands, Fiji, Finland, France, French Guiana, Gabon, Georgia, Germany, Ghana, Greece, Greenland, Grenada, Guam, Guatemala, Guyana, Haiti, Honduras, Hungary, Hong Kong, Iceland, India, Ireland, Israel, Italy, Jamaica, Japan, Kenya, Laos, Latvia, Lesotho, Liberia, Lithuania, Luxembourg, Macedonia, Madagascar, Malawi, Martinique, Mauritius, Mexico, Micronesia, Moldova, Mongolia, Mozambique, Myanmar, Namibia, Netherlands Antilles, Nepal, Netherlands, New Caledonia, New Zealand, Nicaragua, North Korea, Norway, Panama, Papua New Guinea, Paraguay, Peru, Philippines, Poland, Portugal, Puerto Rico, Rep.\ of Congo, Reunion, Romania, Russia, Rwanda, Samoa, Sao Tome and Principe, Serbia, Slovakia, Slovenia, Solomon Islands, South Africa, South Korea, Spain, Sri Lanka, Suriname, Swaziland, Sweden, Switzerland, Taiwan, Thailand, Togo, Tonga, Trinidad and Tobago, Uganda, Ukraine, United Kingdom, United States, Uruguay, Venezuela, Vietnam, Wallis and Futuna, Zambia, Zimbabwe. 

\paragraph{Countries and territories with a Muslim population share between 25\% and 75\%:} 
Albania, Bosnia-Herzegovina, Brunei, Burkina Faso, Chad, C\^{o}te d'Ivoire, Eritrea, Ethiopia, Guinea, Guinea-Bissau, Kazakhstan, Kyrgyzstan, Lebanon, Malaysia, Nigeria, Sierra Leone, Sudan, Tanzania. 

\paragraph{Countries and territories with a Muslim population share above 75\%:} 
Afghanistan, Algeria, Azerbaijan, Bahrain, Bangladesh, Comoros, Djibouti, Egypt, Gambia, Indonesia, Iran, Iraq, Jordan, Kosovo, Kuwait, Libya, Mali, Mauritania, Morocco, Niger, Oman, Pakistan, Palestine, Saudi Arabia, Senegal, Somalia, Syria, Tajikistan, Tunisia, Turkmenistan, Turkey, United Arab Emirates, Uzbekistan, Western Sahara, Yemen.

\bigskip

\begin{figure}[h!]
\caption*{Figure A.1: Country-level population shares of Muslims} \label{fig:map}
\centering
\fbox{\includegraphics[width=0.65\textwidth]{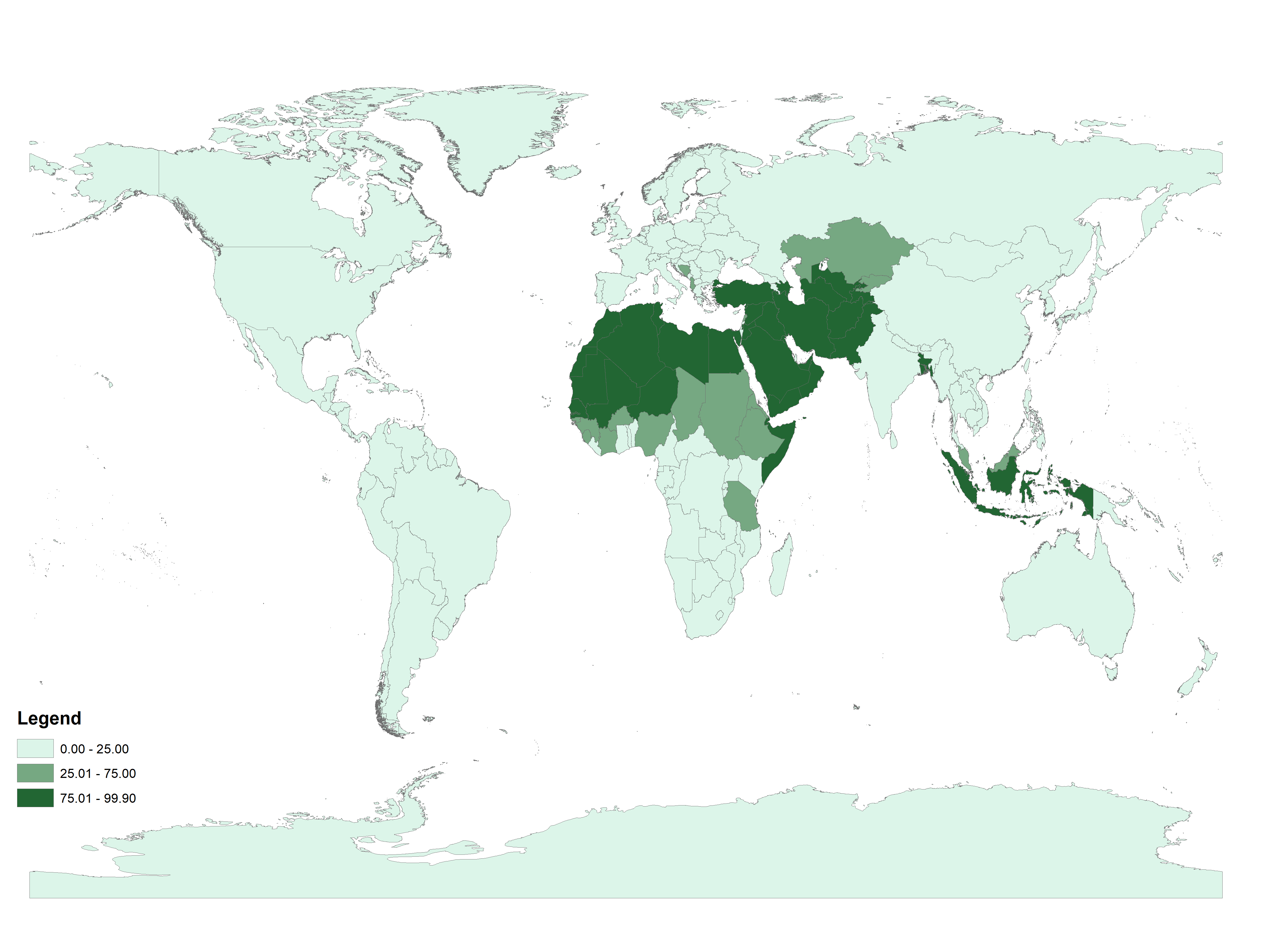} }
\parbox{\textwidth}{\footnotesize \emph{Notes:} Data on Muslim population share (in \%) in 1990 is from the Pew Research Center. Darker colors indicate higher shares.}
\end{figure}

\bigskip

\begin{figure}[h!]
\caption*{Figure A.2: Histogram of Muslim population shares across countries and territories} \label{fig:hist}
\centering
\includegraphics[width=0.7\textwidth]{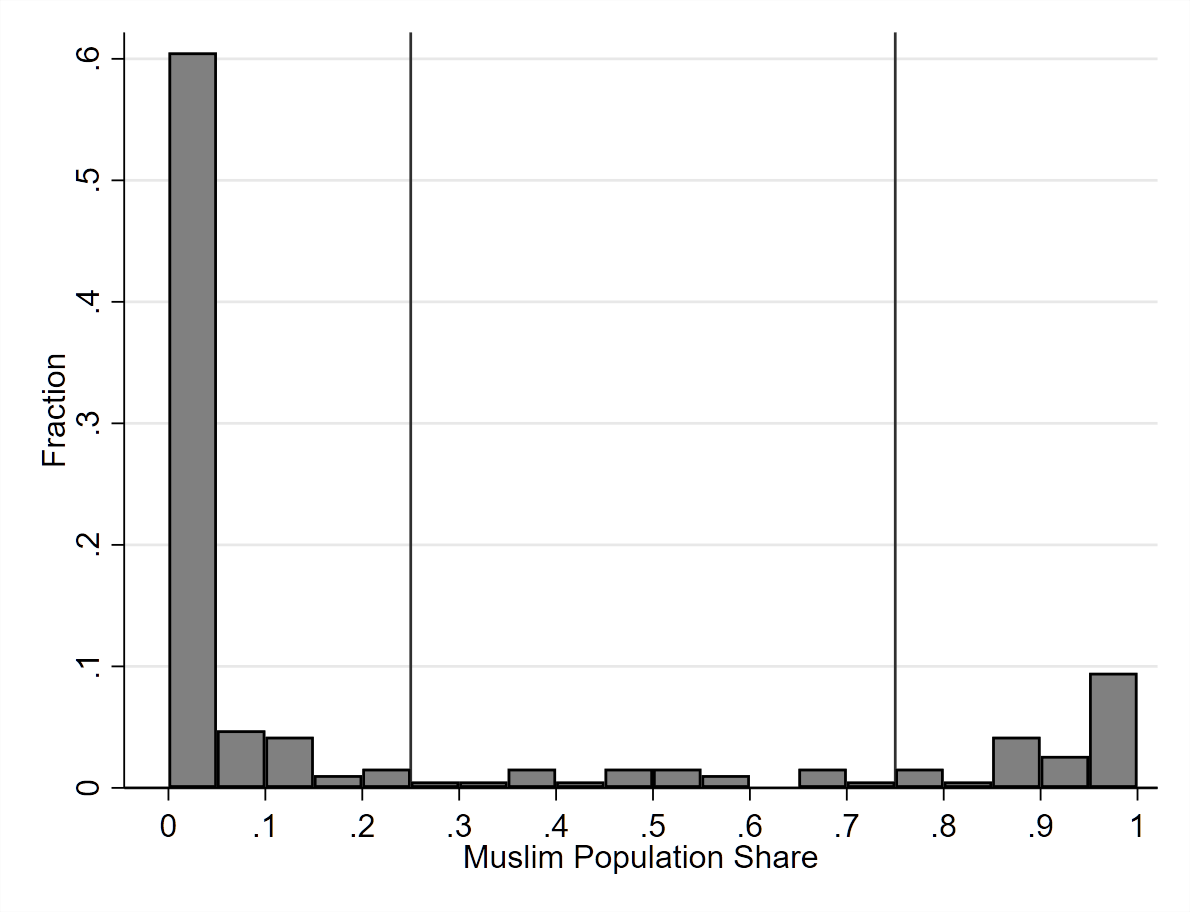} 
\parbox{0.9\textwidth}{\footnotesize \emph{Notes:} Data on Muslim population share in 1990 is from the Pew Research Center.}
\end{figure}

%%%%%%%%%%%%%%%%%%%%%%%%%%%%%%%%%%%%%%%%%%%%%%%%%%%%%%%%%%%%%%%%%%%%%%%%%%%%%%%
\clearpage
\section*{B. Additional descriptive statistics}

\begin{table}[h!]
\centering
\caption*{Table B.1: Terrorist events and terror deaths in countries and territories with more than 1,000 events during sample period} 
    \label{tabD} 
    \begin{tabular}{rlccc} \hline \hline
    Rank & Country & Terrorist events & Terror deaths & Death rate \\ 
	\hline
    1.    & Iraq  & 17,610 & 54,305 & 3.08 \\
    2.    & Pakistan & 11,304 & 17,876 & 1.58 \\
    3.    & India & 8,735 & 14,498 & 1.66 \\
    4.    & Afghanistan & 7,995 & 22,946 & 2.87 \\
    5.    & Colombia & 6,082 & 8,909 & 1.46 \\
    6.    & Peru  & 5,247 & 7,682 & 1.46 \\
    7.    & United Kingdom & 3,933 & 2,475 & 0.63 \\
    8.    & Philippines & 3,735 & 4,114 & 1.10 \\
    9.    & El Salvador & 3,052 & 3,731 & 1.22 \\
    10.   & Spain & 2,854 & 1,117 & 0.39 \\
    11.   & Turkey & 2,729 & 3,131 & 1.15 \\
    12.   & Nigeria & 2,663 & 17,730 & 6.66 \\
    13.   & Thailand & 2,464 & 1,722 & 0.70 \\
    14.   & Chile & 2,222 & 215   & 0.10 \\
    15.   & Sri Lanka & 2,071 & 9,464 & 4.57 \\
    16.   & France & 2,057 & 397   & 0.19 \\
    17.   & Somalia & 1,993 & 4,045 & 2.03 \\
    18.   & Algeria & 1,990 & 8,287 & 4.16 \\
    19.   & Russia & 1,773 & 3,555 & 2.01 \\
    20.   & United States & 1,743 & 3,447 & 1.98 \\
    21.   & Yemen & 1,644 & 3,638 & 2.21 \\
    22.   & Israel & 1,629 & 1,388 & 0.85 \\
    23.   & South Africa & 1,613 & 1,886 & 1.17 \\
    24.   & Lebanon & 1,541 & 2,681 & 1.74 \\
    25.   & Palestina & 1,523 & 1,111 & 0.73 \\
    26.   & Egypt & 1,468 & 2,010 & 1.37 \\
    27.   & Italy & 1,458 & 399   & 0.27 \\
    28.   & Libya & 1,441 & 1,278 & 0.89 \\
    29.   & Bangladesh & 1,370 & 867   & 0.63 \\
    30.   & Guatemala & 1,292 & 2,006 & 1.55 \\
    31.   & Syria & 1,181 & 7,491 & 6.34 \\
    32.   & Greece & 1,134 & 237   & 0.21 \\
    33.   & Germany & 1,052 & 129   & 0.12 \\
%    34.   & Nepal & 853   & 1,489 & 1.75 \\
%    35.   & Nicaragua & 763   & 1,751 & 2.29 \\
%    36.   & Ukraine & 757   & 1,028 & 1.36 \\
%    37.   & Argentina & 689   & 291   & 0.42 \\
%    38.   & Sudan & 682   & 3,486 & 5.11 \\
%    39.   & Indonesia & 543   & 794   & 1.46 \\
%    40.   & Iran  & 525   & 1,464 & 2.79 \\
%    41.   & Dem.\ Rep.\ of Congo & 463   & 2,510 & 5.42 \\
%    42.   & Burundi & 459   & 2,604 & 5.67 \\
%    43.   & Kenya & 447   & 1,512 & 3.38 \\
%    44.   & Mexico & 395   & 526   & 1.33 \\
%    45.   & Angola & 362   & 1,741 & 4.81 \\
%    46.   & Japan & 327   & 41    & 0.13 \\
%    47.   & Uganda & 310   & 2,418 & 7.80 \\
%    48.   & Bolivia & 287   & 24    & 0.08 \\
%    49.   & Myanmar & 272   & 445   & 1.64 \\
%    50.   & Honduras & 270   & 178   & 0.66 \\		
    \hline \hline
    \end{tabular}
    \parbox{\textwidth}{\footnotesize \emph{Notes:} Sample period covers Ramadan years 1390--1436 (corresponding to Gregorian years 1970--2015). Data source is the Global Terror Database. Death rate is the average number of terror deaths per terrorist event.}
\end{table}

\newpage

\begin{table}[h!]
\caption*{Table B.2: Terrorist events and terror deaths worldwide, by Muslim population shares} \label{tab:des_gtd}
\centering
    \begin{tabular}{lccccccc} \hline \hline
          & \multicolumn{6}{c}{Muslim population share} & All \\
          & \multicolumn{2}{c}{$<$25\%} & \multicolumn{2}{c}{25--75\%} & \multicolumn{2}{c}{$>$75\%} &  \\
          & Freq. & Perc. & Freq. & Perc. & Freq. & Perc. & Freq. \\
    \cline{2-8}      & (1)   & (2)   & (3)   & (4)   & (5)   & (6)   & (7) \\ \hline
    \multicolumn{8}{c}{\textbf{Population}} \\\hline
    Population (in mio.) & 4,314  & 79\%  & 313   & 6\%   & 835   & 15\%  & 5,462 \\\hline
    \multicolumn{8}{c}{\textbf{Terrorist events}} \\\hline
    Total & 66,103 & 52\%  & 5,611  & 4\%   & 54,632 & 43\%  & 126,346 \\
	By perpetrators: &&&&&&&\\
   \quad Non-Muslim & 36,154 & 96\%  & 373   & 1\%   & 1,307  & 3\%   & 37,834 \\
   \quad  Muslim & 4,524  & 17\%  & 2,759  & 10\%  & 19,193 & 72\%  & 26,476 \\
   %\quad Islamist & 2,287  & 12\%  & 1,888  & 10\%  & 14,523 & 78\%  & 18,698 \\
   %\quad  Secular Muslim & 1,053  & 28\%  & 417   & 11\%  & 2,308  & 61\%  & 3,778 \\
	By targets: &&&&&&&\\
   \quad Armed targets & 11,493 & 42\%  & 808   & 3\%   & 14,792 & 55\%  & 27,093 \\
   \quad Civilians & 44,169 & 54\%  & 4,211  & 5\%   & 33,830 & 41\%  & 82,210 \\
   \quad Infrastructure & 8,854  & 70\%  & 349   & 3\%   & 3,532  & 28\%  & 12,735 \\
	By attack type: &&&&&&&\\
  \quad  Assault & 13,176 & 50\%  & 1,851  & 7\%   & 11,282 & 43\%  & 26,309 \\
  %\quad  Assassination & 8,861  & 61\%  & 476   & 3\%   & 5,095  & 35\%  & 14,432 \\
  \quad  Bombing & 31,574 & 48\%  & 2,097  & 3\%   & 31,639 & 48\%  & 65,310 \\
  \quad	 Suicide attack & 385   & 9\%   & 269   & 6\%   & 3,542  & 84\%  & 4,196 \\\hline
   \multicolumn{8}{c}{\textbf{Terror deaths}} \\\hline
    Total & 94,603 & 37\%  & 26,997 & 11\%  & 132,508 & 52\%  & 254,108 \\
		By perpetrators: &&&&&&&\\
   \quad   Non-Muslim & 58,255 & 92\%  & 3,131  & 5\%   & 1,941  & 3\%   & 63,327 \\
   \quad   Muslim & 15,502 & 15\%  & 18,713 & 18\%  & 72,478 & 68\%  & 106,693 \\
   %\quad   Islamist & 10,634 & 12\%  & 14,642 & 16\%  & 64,367 & 72\%  & 89,643 \\
   %\quad   Secular Muslim & 1,605  & 24\%  & 1,636  & 25\%  & 3,388  & 51\%  & 6,629 \\
		By targets: &&&&&&&\\
   \quad   Armed targets & 18,371 & 31\%  & 3,071  & 5\%   & 38,672 & 64\%  & 60,114 \\
   \quad  Civilians & 53,294 & 36\%  & 19,613 & 13\%  & 76,890 & 51\%  & 149,797 \\
   \quad    Infrastructure & 8,781  & 59\%  & 864   & 6\%   & 5,277  & 35\%  & 14,922 \\
		By attack type: &&&&&&&\\
  \quad    Assault & 42,534 & 48\%  & 14,675 & 17\%  & 30,990 & 35\%  & 88,199 \\
  %\quad   Assassination & 11,494 & 61\%  & 605   & 3\%   & 6,754  & 36\%  & 18,853 \\
  \quad    Bombing & 24,153 & 22\%  & 7,574  & 7\%   & 77,938 & 71\%  & 109,665 \\
  \quad    Suicide attack & 6,910  & 15\%  & 2,924  & 6\%   & 35,836 & 78\%  & 45,670 \\ \hline\hline
    \end{tabular}
\parbox{\textwidth}{\footnotesize \emph{Notes:} Sample period covers Ramadan years 1390--1436 (corresponding to Gregorian years 1970--2015). Data source for terrorist events and terror deaths is Global Terror Database. Population size is from the World Bank Development Indicator (WBDI) data and averaged over the sample period, but excludes countries and years with missing information. The WBDI does not provide population data for French Guiana, Martinique, Netherlands Antilles, Reunion, Taiwan, Wallis and Futuna, and Western Sahara. Section \ref{sec:data.gtd} describes the terrorism data.}
\end{table}

\clearpage
\newpage 

\begin{table}[h] \centering
\caption*{Table B.3: Daylight hours in Muslim countries and territories} \label{tab:day_hours}
    \begin{tabular}{lccccc}
\hline\hline
     Country        & Obs. & Mean  & Std. Dev. & Max.~Diff. & Max.~Diff. \\
            & & (in hours) & (in hours) & across years & within year \\
            &  & & & (in hours) & (in min.) \\
    \cline{2-6}      & (1)   & (2)   & (3)   & (4)   & (5)  \\\hline
    Afghanistan & 15,416 & 12.3  & 1.5   & 5.2  &43 \\
    Algeria & 70,688 & 12.3  & 1.6   & 5.0  &78 \\
    Azerbaijan & 3,290  & 12.3  & 1.9   & 6.0  &21 \\
    Bahrain & 235   & 12.2  & 1.1   & 3.3  &2 \\
    Bangladesh & 1,081  & 12.2  & 1.0   & 3.2  & 17 \\
    Comoros & 141   & 12.1  & 0.5   & 1.4  & 3 \\
    Djibouti & 517   & 12.2  & 0.5   & 1.5  & 6 \\
    Egypt & 1,222  & 12.3  & 1.2   & 4.0  & 36 \\
    Gambia & 611   & 12.2  & 0.5   & 1.6  & 3 \\
    Indonesia & 18,518 & 12.1  & 0.2   & 1.2  & 59 \\
    Iran  & 12,596 & 12.3  & 1.5   & 5.5 & 72 \\
    Iraq  & 4,794  & 12.3  & 1.5   & 5.0  &37 \\
    Jordan & 2,444  & 12.3  & 1.4   & 4.2  & 17 \\
    Kosovo & 1,410  & 12.4  & 2.0   & 6.3  & 10 \\
    Kuwait & 235   & 12.3  & 1.2   & 3.7 & 3 \\
    Libya & 1,504  & 12.3  & 1.3   & 4.3  & 42 \\
    Mali  & 2,397  & 12.2  & 0.6   & 2.5  & 41 \\
    Mauritania & 2,068  & 12.2  & 0.7   & 3.3 & 46  \\
    Morocco & 2,538  & 12.3  & 1.4   & 4.8 & 40 \\
    Niger & 1,692  & 12.2  & 0.6   & 2.5  &34 \\
    Oman  & 2,303  & 12.2  & 0.9   & 3.3  & 40 \\
    Pakistan & 1,504  & 12.3  & 1.3   & 4.8 & 52 \\
    Palestina & 752   & 12.3  & 1.4   & 4.2  &6 \\
    Saudi Arabia & 611   & 12.2  & 1.0   & 3.8  & 54 \\
    Senegal & 1,410  & 12.2  & 0.6   & 1.9  & 14 \\
    Somalia & 3,478  & 12.1  & 0.3   & 1.4  & 46 \\
    Syria & 658   & 12.3  & 1.5   & 4.9  & 19 \\
    Tajikistan & 2,726  & 12.3  & 1.8   & 5.7  & 23 \\
    Tunisia & 12,314 & 12.3  & 1.6   & 5.1  & 30\\
    Turkey & 3,431  & 12.3  & 1.8   & 6.0 &34 \\
    Turkmenistan & 235   & 12.3  & 1.8   & 5.8 &24 \\
    United Arab Emirates & 376   & 12.2  & 1.0   & 3.1 &9 \\
    Uzbekistan & 658   & 12.3  & 1.9   & 6.3 & 35 \\
    Western Sahara & 188   & 12.2  & 1.0   & 3.3  & 18 \\
    Yemen & 15,651 & 12.2  & 0.6  & 2.1 & 20 \\\hline
  All &   189,692 & 12.3 & 1.4 & 6.3 & 78 \\    \hline \hline
    \end{tabular}
\parbox{\textwidth}{\footnotesize \emph{Notes:} Descriptive statistics based on our own computation of average daylight hours during Ramadan from 1390--1436 for all districts (ADM2 regions). Column (4) provides the maximum difference between the longest and the shorted Ramadan daylight hours, and column (5) the maximum difference in a single Ramadan year.}
\end{table}

\clearpage
\newpage 

\begin{figure}[h!]
\caption*{Figure B.1: Terrorist events and terror deaths by Ramadan year} \label{fig:dev_aggregate}
\centering
\includegraphics[width=0.7\textwidth]{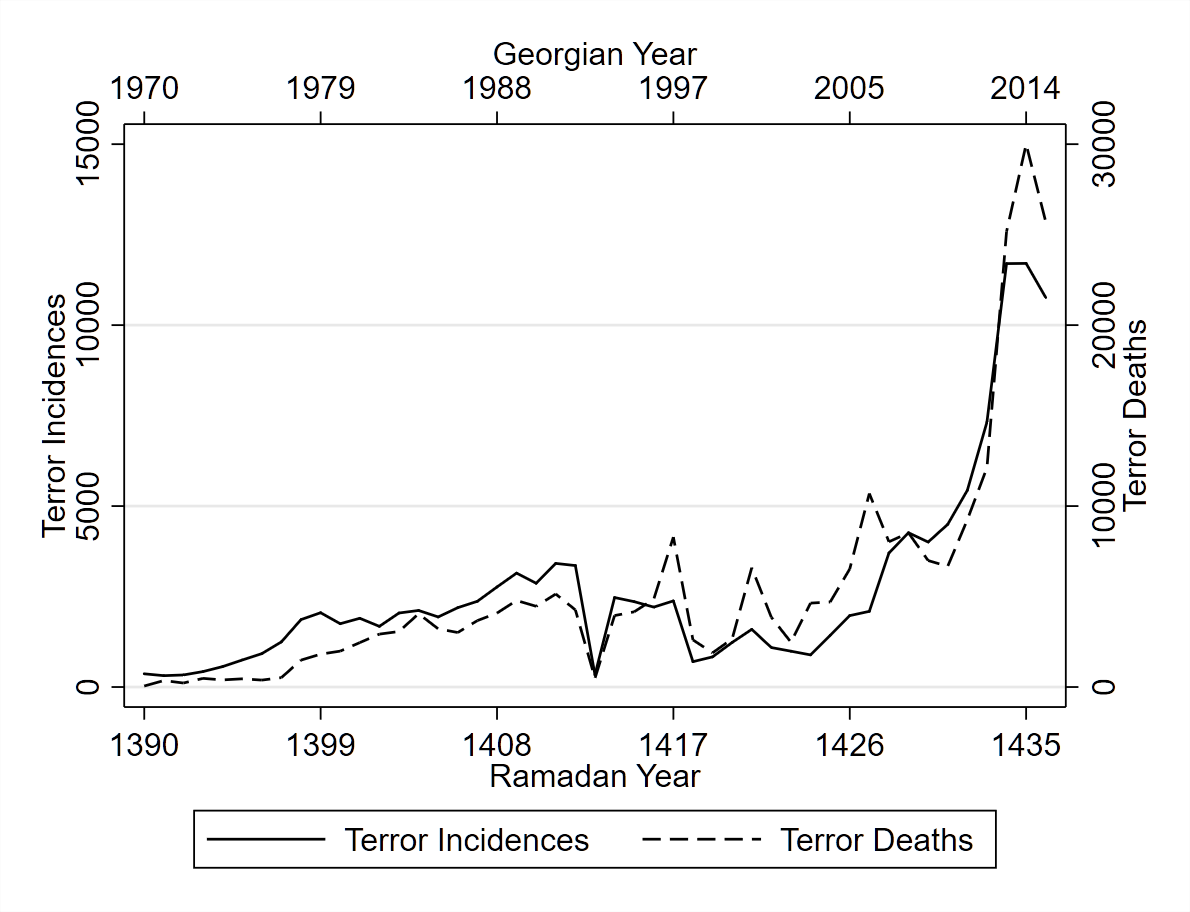} 
\parbox{0.7\textwidth}{\footnotesize \emph{Notes:} Section \ref{sec:data.gtd} describes the terrorism data and section \ref{sec:data.panel} introduces the time unit ``Ramadan year.''}
\end{figure}

\begin{figure}[h!]
\caption*{Figure B.2: Terrorist events and terror deaths by Muslim population share} \label{fig:dev_Mshare}
\centering
\includegraphics[width=0.7\textwidth]{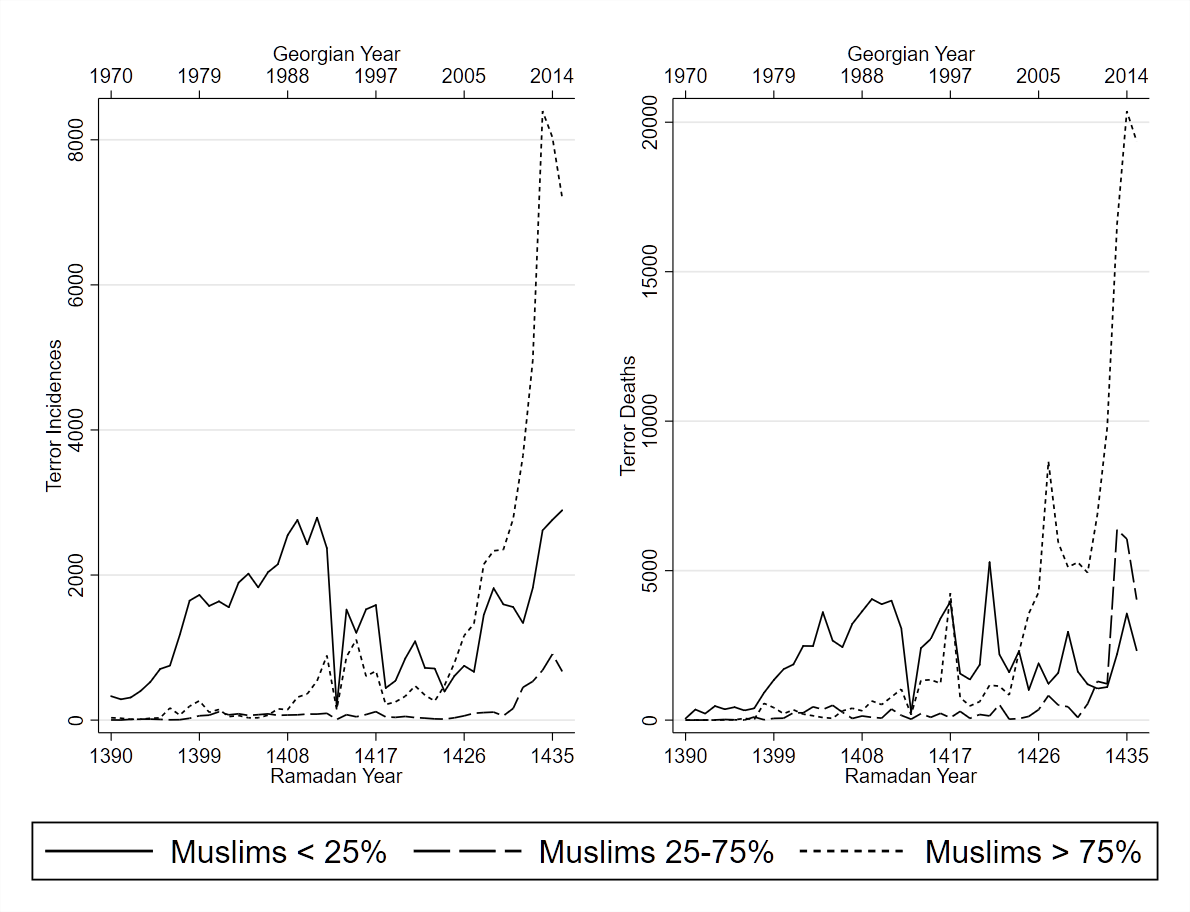} 
\parbox{0.7\textwidth}{\footnotesize \emph{Notes:} Section \ref{sec:data.gtd} describes the terrorism data and section \ref{sec:data.panel} introduces the time unit ``Ramadan year.''}
\end{figure}

\begin{figure}[h!]
\caption*{Figure B.3: Terrorist events and terror deaths by Muslim vs Non-Muslim perpetrator groups} \label{fig:dev_perp}
\centering
\includegraphics[width=0.7\textwidth]{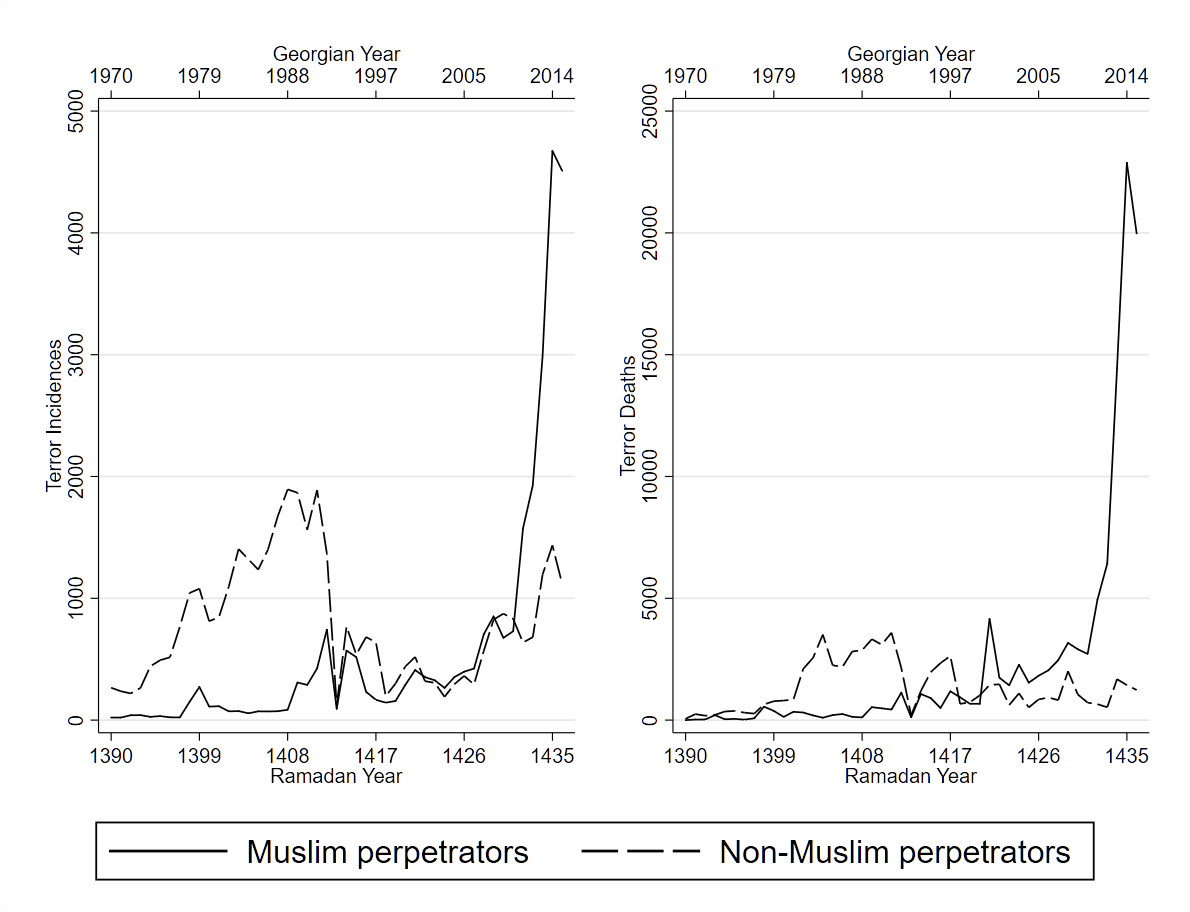} 
\parbox{0.7\textwidth}{\footnotesize \emph{Notes:} Section \ref{sec:data.gtd} describes the terrorism data and section \ref{sec:data.panel} introduces the time unit ``Ramadan year.''}
\end{figure}

%\begin{figure}[h!]
%\caption{Terrorist events and terror deaths of Islamist and secular Muslim perpetrator groups} \label{fig:dev_Mperp}
%\centering
%\includegraphics[width=0.7\textwidth]{fig_dev_Mperp.png} 
%\parbox{0.7\textwidth}{\footnotesize \emph{Notes:} Section \ref{sec:data.gtd} describes the terrorism data and section \ref{sec:data.panel} introduces the time unit ``Ramadan year.''}
%\end{figure}

\begin{figure}[h!]
\caption*{Figure B.4: Terrorist events and terror deaths by targets} \label{fig:dev_targets}
\centering
\includegraphics[width=0.7\textwidth]{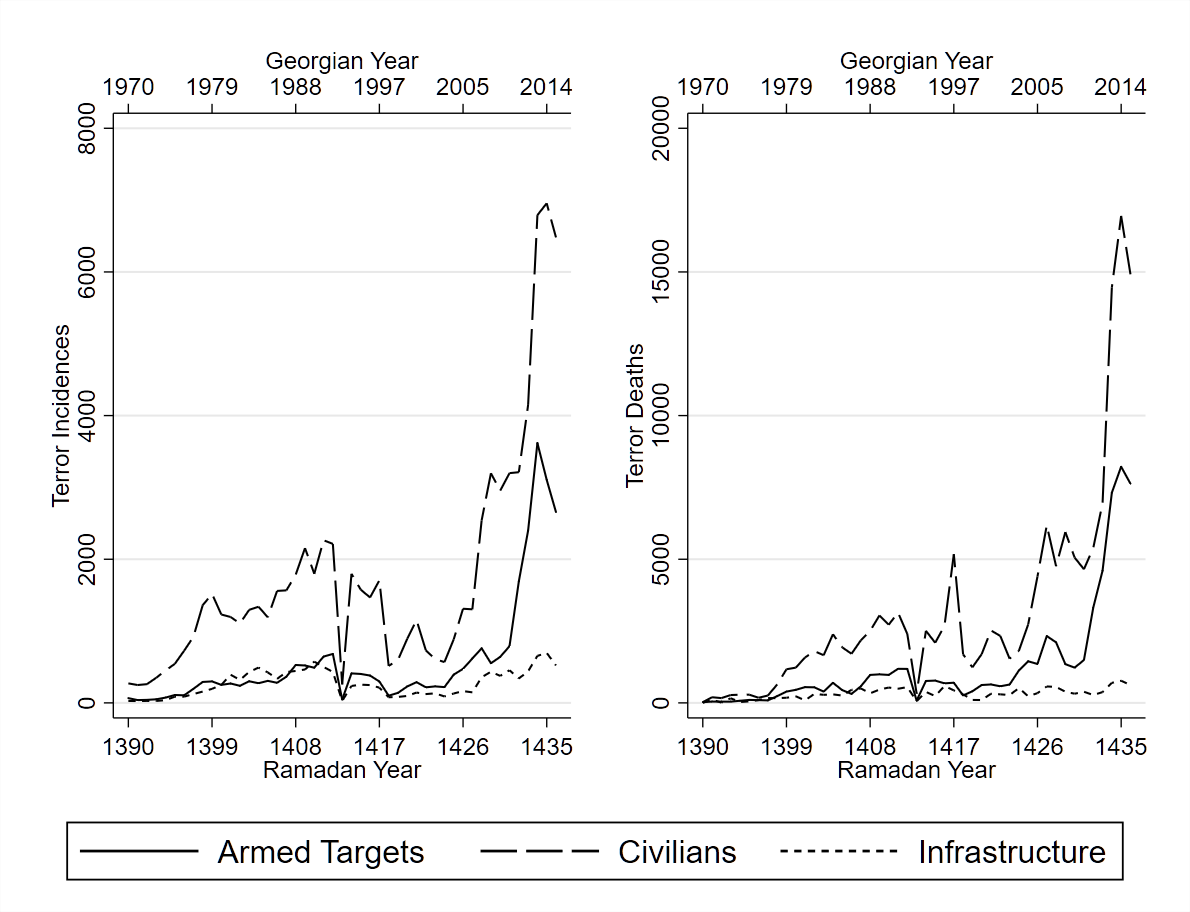} 
\parbox{0.7\textwidth}{\footnotesize \emph{Notes:} Section \ref{sec:data.gtd} describes the terrorism data and section \ref{sec:data.panel} introduces the time unit ``Ramadan year.''}
\end{figure}

\begin{figure}[h!]
\caption*{Figure B.5: Terrorist events and terror deaths by attack types} \label{fig:dev_tactics}
\centering
\includegraphics[width=0.7\textwidth]{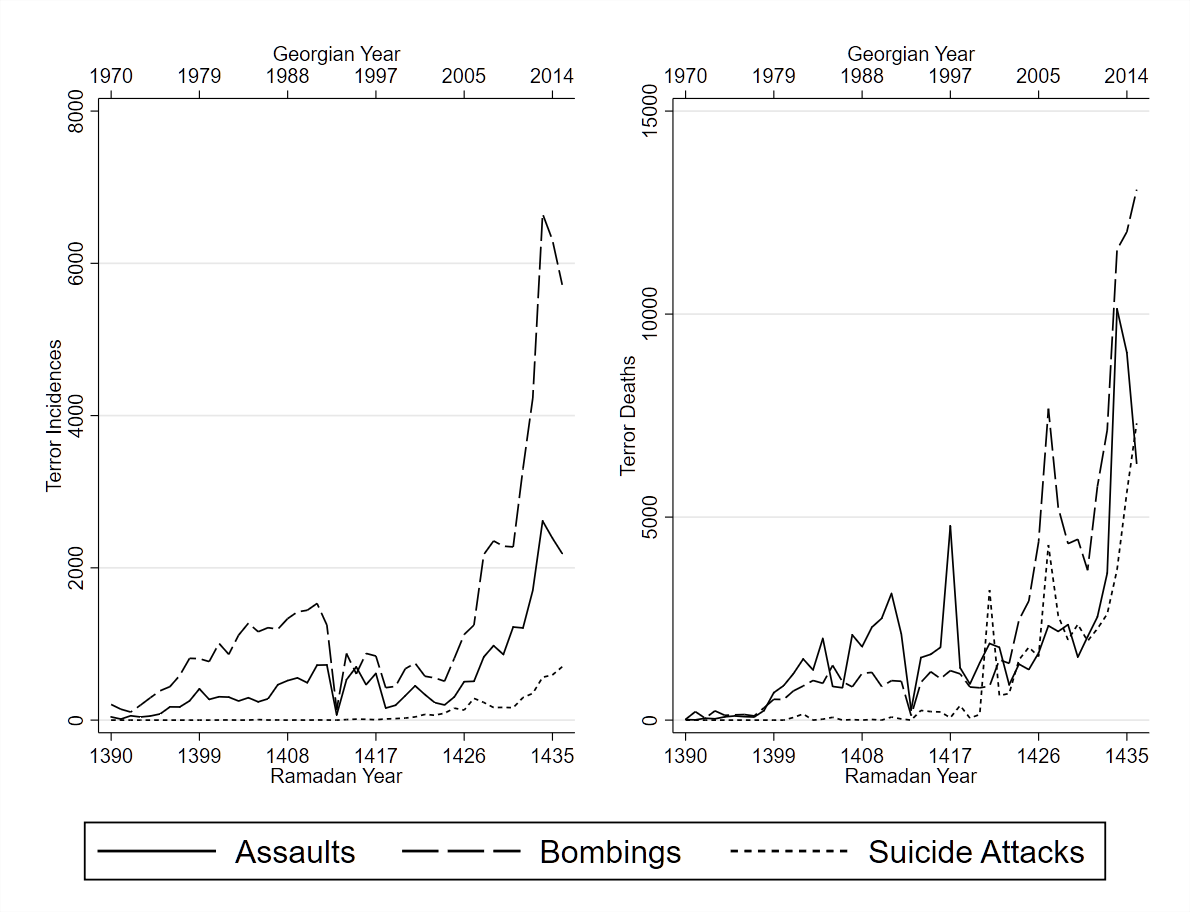} 
\parbox{0.7\textwidth}{\footnotesize \emph{Notes:} Section \ref{sec:data.gtd} describes the terrorism data and section \ref{sec:data.panel} introduces the time unit ``Ramadan year.''}
\end{figure}

\begin{figure}[h!]
\caption*{Figure B.6: Share of terrorist events and terror deaths in the month of Ramadan} \label{fig:duringramadan}
\centering
\includegraphics[width=0.65\textwidth]{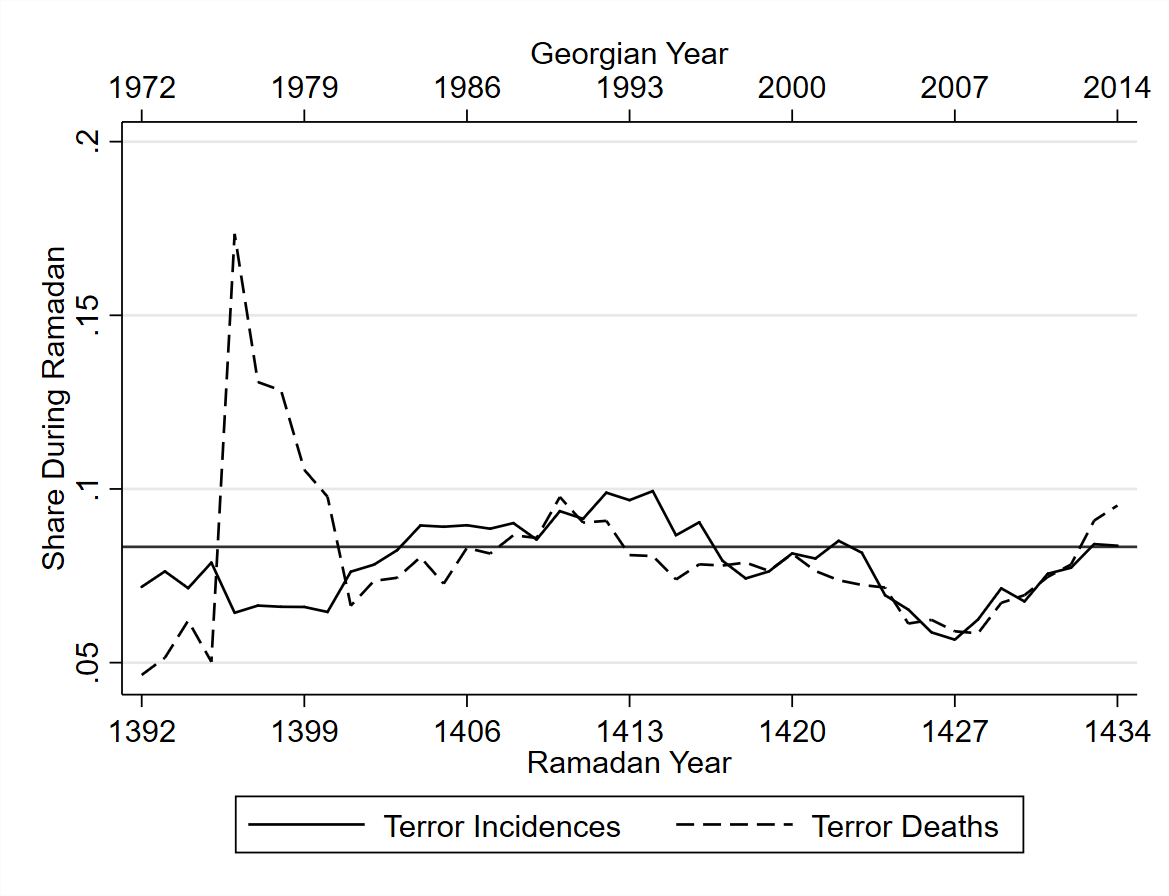} 
\parbox{0.7\textwidth}{\footnotesize \emph{Notes:} The shares are calculated using five year moving averages of the terrorist events and terror deaths. There is a vertical line at a share of 0.833, because the share of terrorist events in Ramadan would be 8.33\% if they were uniformly distributed across the 12 Islamic months. Section \ref{sec:data.gtd} describes the terrorism data and section \ref{sec:data.panel} introduces the time unit ``Ramadan year.''}
\end{figure}

\begin{figure}[h!]
\caption*{Figure B.7: Share of terrorist events and terror deaths committed by Muslim perpetrator groups in the month of Ramadan} \label{fig:duringramadan_mperpetrators}
\centering
\includegraphics[width=0.6\textwidth]{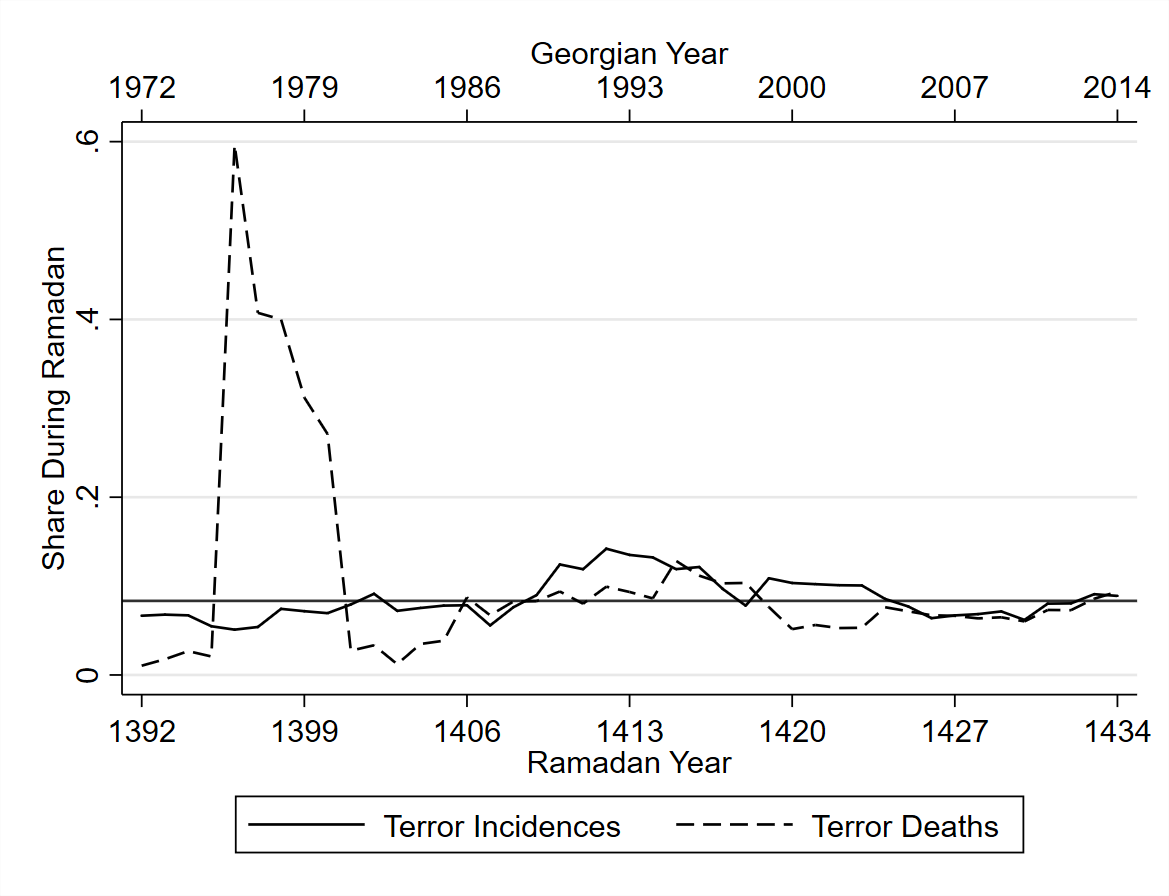} 
\parbox{0.7\textwidth}{\footnotesize \emph{Notes:} The shares are calculated using five year moving averages of the terrorist events and terror deaths. There is a vertical line at a share of 0.833, because the share of terrorist events in Ramadan would be 8.33\% if they were uniformly distributed across the 12 Islamic months. Section \ref{sec:data.gtd} describes the terrorism data and section \ref{sec:data.panel} introduces the time unit ``Ramadan year.''}
\end{figure}

\begin{figure}[h!]
\caption*{Figure B.8: Share of terrorist events and terror deaths in the month of Ramadan in countries with a Muslim population share above 75\%} \label{fig:duringramadan_mcountries}
\centering
\includegraphics[width=0.6\textwidth]{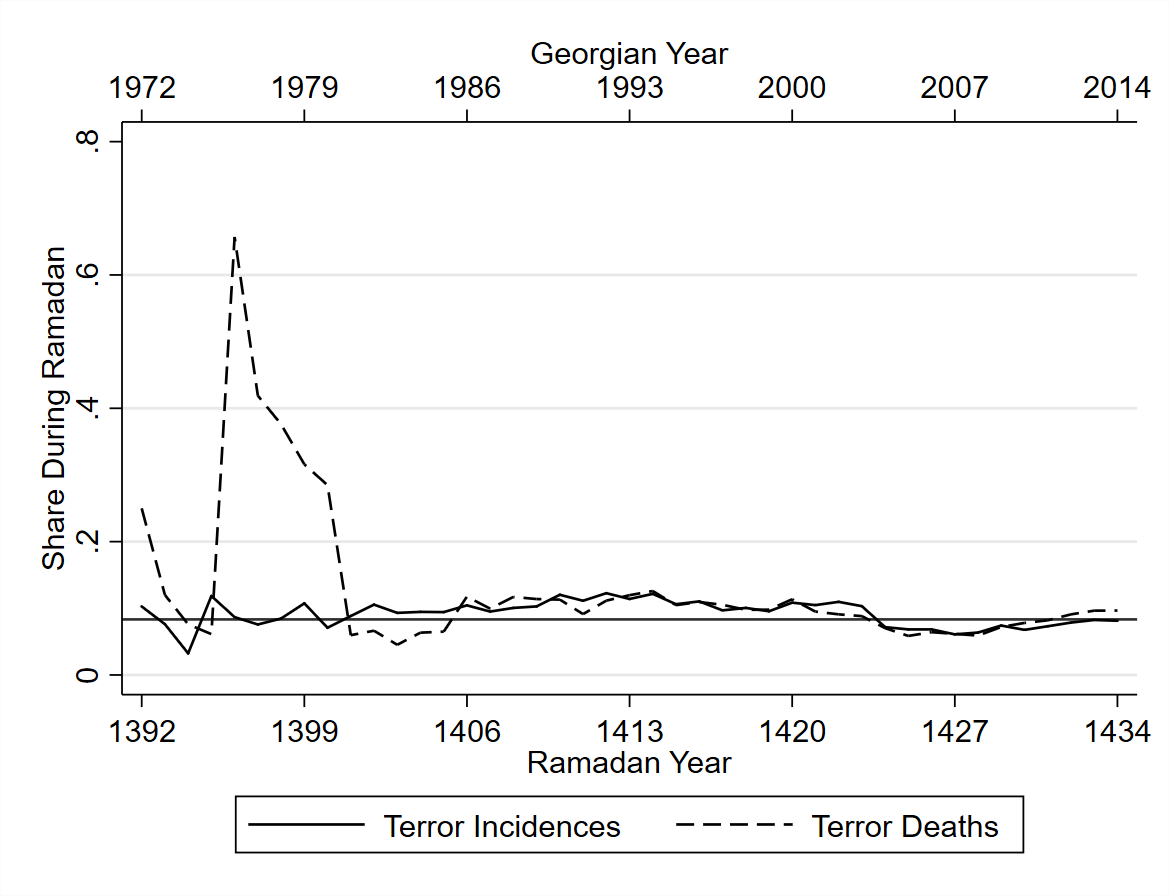} 
\parbox{0.7\textwidth}{\footnotesize \emph{Notes:} The shares are calculated using five year moving averages of the terrorist events and terror deaths. There is a vertical line at a share of 0.833, because the share of terrorist events in Ramadan would be 8.33\% if they were uniformly distributed across the 12 Islamic months. Section \ref{sec:data.gtd} describes the terrorism data and section \ref{sec:data.panel} introduces the time unit ``Ramadan year.''}
\end{figure}

%%%%%%%%%%%%%%%%%%%%%%%%%%%%%%%%%%%%%%%%%%%%%%%%%%%%%%%%%%%%%%%%%%%%%%%%%%%%%%%
\clearpage
\newpage 
\section*{C. Robustness tests}

\begin{table}[h!]  
\caption*{Table C.1: Effects of Ramadan daylight hours on the number of terrorist events and terror deaths} \label{tab:intensive}\centering
\begin{tabular}{lcccc}\hline\hline
 & (1)   & (2)   & (3)   & (4) \\\hline
\multicolumn{5}{c}{A: Terrorist events (log-modulus)} \\\hline
Ramadan daylight hours  &      -0.071   &      -0.075   &       0.026   &               \\
            &     (0.101)   &     (0.105)   &     (0.124)   &               \\
\quad $\times$ Muslim pop.\ share    &               &               &      -1.931   &               \\
            &               &               &     (1.656)   &               \\
\quad $\times$ Muslims $<$25\%  &               &               &               &      -0.057   \\
            &               &               &               &     (0.097)   \\
\quad $\times$ Muslims 25-75\%  &               &               &               &       2.071   \\
            &               &               &               &     (2.126)   \\
\quad $\times$ Muslims $>$75\%  &               &               &               &      -3.286** \\
            &               &               &               &     (1.582)   \\
\hline
\multicolumn{5}{c}{B: Terror deaths (log-modulus)} \\\hline
Ramadan daylight hours  &      -0.052   &      -0.024   &       0.105   &               \\
            &     (0.087)   &     (0.126)   &     (0.157)   &               \\
\quad $\times$ Muslim pop.\ share    &               &               &      -2.474   &               \\
            &               &               &     (2.732)   &               \\
\quad $\times$ Muslims $<$25\%  &               &               &               &      -0.010   \\
            &               &               &               &     (0.111)   \\
\quad $\times$ Muslims 25-75\%  &               &               &               &       4.488   \\
            &               &               &               &     (4.941)   \\
\quad $\times$ Muslims $>$75\%  &               &               &               &      -4.646** \\
            &               &               &               &     (2.336)   \\
\hline
Country fixed effects (FE) & Yes & No & No & No \\
Ramadan year FE & Yes & No & No & No \\
District FE & No & Yes & Yes & Yes \\
Country-Ramadan year FE & No & Yes & Yes & Yes \\
Observations & 1,874,642 & 1,874,642 & 1,874,642 & 1,874,642 \\ \hline\hline
\end{tabular}
\parbox{\textwidth}{\footnotesize \emph{Notes:} Panel fixed effects regressions. Dependent variables are the log modulus of terrorist events in panel A and for terror deaths in panel B, with the log-modulus being multiplied by 100. Section \ref{sec:data} introduces the time unit ``Ramadan year'' and all the variables used. Standard errors are clustered at the country-level and reported in parentheses. ***/**/* indicate statistical significance at the 1\%/5\%/10\%-level.}
\end{table} 

\newpage%%%%%%%%%%%%%%%%%%%%%%%%%%%%%%%%%%%%%%%%%%%%%%%%%%%%%%

\begin{table}[h!] 
\caption*{Table C.2: Non-linear effects of Ramadan daylight hours, by country groups} \label{tab:nonlinear}\centering
\begin{small}
\begin{tabular}{lcccccc}\hline\hline
 & \multicolumn{6}{c}{Muslim population shares:} \\
 & $<$25\% & 25-75\% & $>$75\% & $<$25\% & 25-75\% & $>$75\% \\
\cline{2-7}
  & (1)   & (2)   & (3)  & (4)   & (5)   & (6)  \\
\hline 
 & \multicolumn{3}{c}{Terrorist event (ext. margin)} & \multicolumn{3}{c}{Terror deaths (ext. margin)} \\
\cline{2-4} \cline{5-7} 
Ramadan daylight $<$10.5h 	&       0.251   &       0.006   &      -0.115   &       0.259   &      -0.093   &      -0.234   \\
            &     (0.223)   &     (0.578)   &     (1.402)   &     (0.174)   &     (0.569)   &     (1.272)   \\ 
Ramadan daylight 10.5-11.5h	&       0.119   &      -0.064   &      -0.442   &       0.095   &       0.046   &      -0.430   \\
            &     (0.198)   &     (0.409)   &     (1.041)   &     (0.166)   &     (0.349)   &     (0.976)   \\ 
Ramadan daylight 12.5-13.5h	&      -0.525   &       2.221***&      -2.195** &      -0.261   &       2.386***&      -1.972** \\
            &     (0.393)   &     (0.725)   &     (1.039)   &     (0.194)   &     (0.737)   &     (0.888)   \\
Ramadan daylight $>$13.5	&      -0.706   &       2.389***&      -2.983** &      -0.342   &       2.588***&      -2.668** \\
            &     (0.435)   &     (0.743)   &     (1.346)   &     (0.227)   &     (0.750)   &     (1.208)   \\  
\hline
District fixed effects (FE) & Yes & Yes & Yes & Yes & Yes & Yes \\
Country-Ramadan year FE & Yes & Yes & Yes & Yes & Yes & Yes \\
Observations & 1,591,279 & 93,671 & 189,692 & 1,591,279 & 93,671 & 189,692 \\ \hline\hline
\end{tabular}
\end{small}
\parbox{\textwidth}{\footnotesize \emph{Notes:} Panel regressions with district and country-Ramadan year fixed effects. Sample is restricted to countries with a Muslim population share below 25\% in columns (1) and (4); between 25\% and 75\% in columns (2) and (5); and above 75\% in columns (3) and (6). Dependent variables are the extensive margins for terrorist events in columns (1)--(3) and terror deaths in columns (4)--(6). The explanatory variable are indicator variables for the Ramadan daylight hours. The omitted category are Ramadan daylight hours between 11.5h and 12.5h. Section \ref{sec:data} introduces the time unit ``Ramadan year'' and all the variables used. Standard errors are clustered at the country-level and reported in parentheses. ***/**/* indicate statistical significance at the 1\%/5\%/10\%-level.}
\end{table}

\newpage%%%%%%%%%%%%%%%%%%%%%%%%%%%%%%%%%%%%%%%%%%%%%%%%%%%%%%

%\begin{landscape}
\begin{table}[h!] 
\caption*{Table C.3: Different thresholds for the Muslim population share} \label{tab:thresholds}\centering
\begin{footnotesize}
\begin{tabular}{lcccccc}\hline\hline
 & \multicolumn{3}{c}{Terrorist event} & \multicolumn{3}{c}{Terror deaths} \\
 & \multicolumn{3}{c}{(extensive margin)} & \multicolumn{3}{c}{(extensive margin)} \\\cline{2-7}
 & (1)   & (2)   & (3)  & (4)   & (5)   & (6)  \\\hline
Ramadan daylight hours  & & & & \\
\quad $\times$ Muslims $<$10\%&      -0.036   &               &               &      -0.008   &               &               \\
            &     (0.071)   &               &               &     (0.036)   &               &               \\
\quad $\times$ Muslims 10-90\%&      -0.569   &               &               &      -0.254   &               &               \\
            &     (0.751)   &               &               &     (0.754)   &               &               \\
\quad $\times$ Muslims $>$90\%	&      -2.729*  &               &               &      -2.412** &               &               \\
            &     (1.500)   &               &               &     (1.196)   &               &               \\
\quad $\times$ Muslims $<$20\%&               &      -0.051   &               &               &      -0.019   &               \\
            &               &     (0.076)   &               &               &     (0.040)   &               \\
\quad $\times$ Muslims 20-80\%&               &       1.163   &               &               &       1.316   &               \\
            &               &     (1.202)   &               &               &     (1.360)   &               \\
\quad $\times$ Muslims $>$80\%&               &      -2.733** &               &               &      -2.377** &               \\
            &               &     (1.214)   &               &               &     (0.976)   &               \\
\quad $\times$ Muslims $<$30\%&               &               &      -0.049   &               &               &      -0.018   \\
            &               &               &     (0.075)   &               &               &     (0.040)   \\
\quad $\times$ Muslims 30-70\% &               &               &       1.450   &               &               &       1.651   \\
            &               &               &     (1.307)   &               &               &     (1.518)   \\
\quad $\times$ Muslims $>$70\%&               &               &      -2.942** &               &               &      -2.557***\\
            &               &               &     (1.170)   &               &               &     (0.939)   \\
\hline
District fixed effects (FE) & Yes & Yes & Yes & Yes & Yes & Yes \\
Country-Ramadan year FE & Yes & Yes & Yes & Yes & Yes & Yes \\
Observations & 1,874,642 & 1,874,642 & 1,874,642 & 1,874,642 & 1,874,642 & 1,874,642 \\ \hline\hline
\end{tabular}
\end{footnotesize}
\parbox{\textwidth}{\footnotesize \emph{Notes:} Panel regressions with district and country-Ramadan year fixed effects. Dependent variables are the extensive margins for terrorist events in columns (1)--(3) and terror deaths in columns (4)--(6). Section \ref{sec:data} introduces the time unit ``Ramadan year'' and all the variables used. Standard errors are clustered at the country-level and reported in parentheses. ***/**/* indicate statistical significance at the 1\%/5\%/10\%-level.}
\end{table}
%\end{landscape}

\newpage%%%%%%%%%%%%%%%%%%%%%%%%%%%%%%%%%%%%%%%%%%%%%%%%%%%%%%

\begin{table}[h!] 
\caption*{Table C.4: Country classification according to population share of largest Islamic sect} \label{tab:sects}\centering
\begin{tabular}{lcc}\hline\hline
 & \multicolumn{1}{c}{Terrorist event} & \multicolumn{1}{c}{Terror deaths} \\ 
 & \multicolumn{1}{c}{(extensive margin)} & \multicolumn{1}{c}{(extensive margin)} \\ \cline{2-3}
 & (1)   & (2)  \\\hline
Ramadan daylight hours  & & \\
\quad $\times$ Largest sect $<$25\%  &      -0.049   &      -0.018   \\
            &     (0.075)   &     (0.040)   \\
\quad $\times$ Largest sect 25-75\%  &       0.666   &       0.869   \\
            &     (1.148)   &     (1.305)   \\
\quad $\times$ Largest sect $>$75\%	&      -2.612** &      -2.258** \\
            &     (1.278)   &     (1.030)   \\
\hline
District fixed effects (FE) & Yes & Yes \\
Country-Ramadan year FE & Yes & Yes \\
Observations & 1,874,642 & 1,874,642 \\ \hline\hline
\end{tabular}
\parbox{\textwidth}{\footnotesize \emph{Notes:} Panel regressions with district and country-Ramadan year fixed effects. Dependent variables are the extensive margins for terrorist events in column (1) and terror deaths in column (2). Section \ref{sec:data} describes the time unit ``Ramadan year,'' the dependent variables and the Ramadan daylight hours. Section \ref{sec:results_main} describes the indicator variables that stratify the countries into three groups based on the population share of the largest Islamic sect in any given country. Standard errors are clustered at the country-level and reported in parentheses. ***/**/* indicate statistical significance at the 1\%/5\%/10\%-level.}
\end{table}

\newpage%%%%%%%%%%%%%%%%%%%%%%%%%%%%%%%%%%%%%%%%%%%%%%%%%%%%%%

\begin{table}[h!] 
\caption*{Table C.5: Controls for local population and local economic activity} \label{tab:controls}\centering
\begin{small}
\begin{tabular}{lcccccc}\hline\hline
 & \multicolumn{3}{c}{Terrorist event (ext.\ margin)} & \multicolumn{3}{c}{Terror deaths (ext.\ margin)} \\
\cline{2-4} \cline{5-7} 
  & (1)   & (2)   & (3)  & (4)   & (5)   & (6)  \\
\hline
Ramadan daylight hours  & & & & & & \\
\quad $\times$ Muslims $<$25\%  &      -0.456***&      -0.458***&      -0.459***&      -0.277** &      -0.278** &      -0.278** \\
            &     (0.175)   &     (0.175)   &     (0.175)   &     (0.115)   &     (0.115)   &     (0.115)   \\
\quad $\times$ Muslims 25-75\%  &       1.799   &       1.813   &       1.713   &       2.922   &       2.932   &       2.877   \\
            &     (2.552)   &     (2.564)   &     (2.455)   &     (3.440)   &     (3.448)   &     (3.376)   \\
\quad $\times$ Muslims $>$75\%  &      -5.604** &      -5.599** &      -5.569** &      -5.742** &      -5.738** &      -5.686** \\
            &     (2.233)   &     (2.234)   &     (2.249)   &     (2.420)   &     (2.422)   &     (2.437)   \\
Population (in logs)&               &       0.119*  &               &               &       0.090   &               \\
            &               &     (0.071)   &               &               &     (0.056)   &               \\
\quad $\times$ Muslims $<$25\%  &               &               &       0.102   &               &               &       0.071   \\
            &               &               &     (0.069)   &               &               &     (0.048)   \\
\quad $\times$ Muslims 25-75\%  &               &               &       0.534   &               &               &      -0.598   \\
            &               &               &     (1.027)   &               &               &     (0.965)   \\
\quad $\times$ Muslims $>$75\%  &               &               &       0.663   &               &               &       0.976   \\
            &               &               &     (0.542)   &               &               &     (0.690)   \\
Nighttime lights (in logs)&               &       0.135   &               &               &       0.093   &               \\
            &               &     (0.105)   &               &               &     (0.091)   &               \\
\quad $\times$ Muslims $<$25\%  &               &               &       0.158***&               &               &       0.094** \\
            &               &               &     (0.052)   &               &               &     (0.045)   \\
\quad $\times$ Muslims 25-75\%  &               &               &      -0.598*  &               &               &      -0.441** \\
            &               &               &     (0.322)   &               &               &     (0.208)   \\
\quad $\times$ Muslims $>$75\%  &               &               &       0.444   &               &               &       0.452   \\
            &               &               &     (0.806)   &               &               &     (0.714)   \\
\hline
District fixed effects (FE) & Yes & Yes & Yes & Yes & Yes & Yes \\
Country-Ramadan year FE & Yes & Yes & Yes & Yes & Yes & Yes \\
Observations & 915,763   &  915,763   &  915,763   &  915,763   & 915,763   &  915,763  \\ \hline\hline
\end{tabular}
\end{small}
\parbox{\textwidth}{\footnotesize \emph{Notes:} Panel regressions with district and country-Ramadan year fixed effects. Dependent variables are the extensive margins for terrorist events in columns (1)--(3) and terror deaths in columns (4)--(6). Section \ref{sec:data} introduces the time unit ``Ramadan year'' and most of the variables used.
Population (in logs) is based on the population data from the Center for International Earth Science Information Network (CIESIN). Nighttime lights (in logs) is based on the intensity of nighttime lights measured by weather satellites and provided by the National Oceanic and Atmospheric Administration (NOAA). We add a small constant (0.01) before taking the logarithm of a district's average nighttime light intensity. The sample is restricted to all observations for which the population and nighttime lights data are available. Standard errors are clustered at the country-level and reported in parentheses. ***/**/* indicate statistical significance at the 1\%/5\%/10\%-level.}
\end{table}	

\newpage%%%%%%%%%%%%%%%%%%%%%%%%%%%%%%%%%%%%%%%%%%%%%%%%%%%%%%

\begin{table}[h!] 
\caption*{Table C.6: Provinces, i.e., administrative regions at the first subnational level (ADM1)} \label{tab:adm1}\centering
\begin{tabular}{lcc}\hline\hline
 & \multicolumn{1}{c}{Terrorist event} & \multicolumn{1}{c}{Terror deaths} \\ 
 & \multicolumn{1}{c}{(extensive margin)} & \multicolumn{1}{c}{(extensive margin)} \\ \cline{2-3}
 & (1)   & (2)  \\\hline
Ramadan daylight hours  & & \\
\quad $\times$ Muslims $<$25\% &0.302
&0.194 \\
& (0.537) & (0.481) \\
\quad $\times$ Muslims 25-75\% & 1.389 & 1.119  \\
& (3.743) & (3.283) \\
\quad $\times$ Muslims $>$75\%	& -6.112 & -5.538 \\
& (6.326) & (6.283) \\
\hline
Province fixed effects (FE) & Yes & Yes \\
Country-Ramadan year FE & Yes & Yes \\
Observations & 147,486 & 147,486 \\ \hline\hline
\end{tabular}
\parbox{\textwidth}{\footnotesize \emph{Notes:} Panel regressions with province and country-Ramadan year fixed effects. The sample is based on the aggregation of our standard panel dataset at the province level. Dependent variables are the extensive margins for terrorist events in column (1) and terror deaths in column (2). Section \ref{sec:data} introduces the time unit ``Ramadan year'' and the dependent variables. In our province-level panel dataset, the means [standard deviations] of the dependent variables in columns (1) and (2) are 10.121 [30.161] and 6.048 [23.838], respectively. Standard errors are clustered at the country-level and reported in parentheses. ***/**/* indicate statistical significance at the 1\%/5\%/10\%-level.}
\end{table}

\newpage%%%%%%%%%%%%%%%%%%%%%%%%%%%%%%%%%%%%%%%%%%%%%%%%%%%%%%

\begin{table}[h!] 
\caption*{Table C.7: Different time periods} \label{tab:periods}\centering
\begin{tabular}{lccc}\hline\hline
 & \multicolumn{3}{c}{Ramadan years:} \\
 & $<$1412 & 1412-1422 & $>$1422 \\
\cline{2-4}
  & (1)   & (2)   & (3)  \\
\hline
\multicolumn{4}{c}{A: Terrorist event (extensive margin)}  \\
\hline
Ramadan daylight hours  & & & \\
\quad $\times$ Muslims $<$25\%  &       0.008   &       0.091   &      -0.388**  \\
            &     (0.046)   &     (0.141)   &     (0.179)   \\
\quad $\times$ Muslims 25-75\%  &      -0.162   &       0.422   &       2.763   \\
            &     (0.178)   &     (0.705)   &     (3.650)   \\
\quad $\times$ Muslims $>$75\%  &      -0.142   &      -4.433*  &      -5.095** \\
            &     (0.177)   &     (2.650)   &     (2.344)   \\
\hline
\multicolumn{4}{c}{B: Terror deaths (extensive margin)}  \\
\hline
Ramadan daylight hours  & & & \\
\quad $\times$ Muslims $<$25\%  &       0.013   &       0.057   &      -0.195** \\
            &     (0.038)   &     (0.075)   &     (0.087)   \\
\quad $\times$ Muslims 25-75\%  &      -0.094   &       1.270***&       3.542   \\
            &     (0.112)   &     (0.339)   &     (4.665)   \\
\quad $\times$ Muslims $>$75\%  &      -0.142   &      -4.433*  &      -5.095** \\
            &     (0.145)   &     (2.268)   &     (1.837)   \\
\hline
District fixed effects (FE) & Yes & Yes & Yes \\
Country-Ramadan year FE & Yes & Yes & Yes \\
Observations & 877,492 & 398,860 & 598,290 \\ \hline\hline
\end{tabular}
\parbox{\textwidth}{\footnotesize \emph{Notes:} Panel regressions with district and country-Ramadan year fixed effects. Sample is restricted to different time periods indicated in the top row. (Ramadan years 1412 and 1421 corresponds to Gregorian years 1991 and 2000.) Dependent variables are the extensive margins for terrorist events in panel A and terror deaths in panel B. Section \ref{sec:data} introduces the time unit ``Ramadan year'' and all the variables used. Standard errors are clustered at the country-level and reported in parentheses. ***/**/* indicate statistical significance at the 1\%/5\%/10\%-level.}
\end{table}	

\newpage%%%%%%%%%%%%%%%%%%%%%%%%%%%%%%%%%%%%%%%%%%%%%%%%%%%%%%

\begin{table}[h!]
\caption*{Table C.8: Dropping Ramadan years due to missing data in the GTD} \label{tab:dropGTD}\centering
%\begin{small}
\begin{tabular}{lcc}\hline\hline
 & \multicolumn{2}{c}{Omitted Ramadan years:} \\
 & 1413 & 1412 \& 1413 \\
\cline{2-3}  
 & (1)   & (2)  \\
\hline
\multicolumn{3}{c}{A: Terrorist event (extensive margin)} \\\hline
Ramadan daylight hours  & & \\
\quad $\times$ Muslims $<$25\%  &      -0.047   &      -0.046   \\
            &     (0.076)   &     (0.076)   \\
\quad $\times$ Muslims 25-75\%  &       1.025   &       1.028   \\
            &     (1.205)   &     (1.205)   \\
\quad $\times$ Muslims $>$75\%  &      -2.716** &      -2.710** \\
            &     (1.215)   &     (1.212)   \\
\hline
\multicolumn{3}{c}{B: Terror deaths (extensive margin)} \\\hline
Ramadan daylight hours  & & \\
\quad $\times$ Muslims $<$25\%  &      -0.017   &      -0.016   \\
            &     (0.041)   &     (0.041)   \\
\quad $\times$ Muslims 25-75\%  &       1.233   &       1.238   \\
            &     (1.382)   &     (1.379)   \\
\quad $\times$ Muslims $>$75\%  &      -2.360** &      -2.359** \\
            &     (0.974)   &     (0.969)   \\
\hline
District fixed effects (FE) & Yes & Yes \\
Country-Ramadan year FE & Yes & Yes \\
Observations & 1,834,756   & 1,794,870   \\ \hline\hline
\end{tabular}
%\end{small}
\parbox{\textwidth}{\footnotesize \emph{Notes:} Panel regressions with district and country-Ramadan year fixed effects. Column (1) omits Ramadan year 1413, which started on the Gregorian date February 23, 1993. Column (2) omits Ramadan years 1412 and 1413. The rationale is that the GTD contains no terrorist events for the Gregorian year 1993 ``because they were lost prior to START's compilation of the GTD from multiple data collection efforts'' \citep[][p.\ 3]{start_codebook}. Dependent variables are the extensive margins for terrorist events in panel A and terror deaths in panel B. Section \ref{sec:data} introduces the time unit ``Ramadan year'' and all the variables used. Standard errors are clustered at the country-level and reported in parentheses. ***/**/* indicate statistical significance at the 1\%/5\%/10\%-level.}
\end{table}	

\newpage%%%%%%%%%%%%%%%%%%%%%%%%%%%%%%%%%%%%%%%%%%%%%%%%%%%%%%

\begin{table}[h!]  
\caption*{Table C.9: Including ambiguous terror events} \label{} \centering
\begin{tabular}{lcccc}\hline\hline
 & (1)   & (2)   & (3)   & (4) \\\hline
\multicolumn{5}{c}{A: Terrorist event (extensive margin)} \\\hline
Ramadan daylight hours          &  -0.100   &      -0.065   &       0.037   &               \\
            &     (0.086)   &     (0.100)   &     (0.120)   &               \\
\quad $\times$ Muslim pop.\ share &               &               &      -1.939   &               \\
            &               &               &     (1.598)   &               \\
\quad $\times$ Muslims $<$25\% &               &               &               &      -0.041   \\
            &               &               &               &     (0.090)   \\
\quad $\times$ Muslims 25-75\% &               &               &               &       1.111   \\
            &               &               &               &     (1.283)   \\
\quad $\times$ Muslims $>$75\% &               &               &               &      -3.104** \\
            &               &               &               &     (1.506)   \\
\hline
\multicolumn{5}{c}{B: Terror deaths (extensive margin)} \\\hline
Ramadan daylight hours     &     -0.059   &      -0.038   &       0.059   &               \\
            &     (0.056)   &     (0.066)   &     (0.088)   &               \\
\quad $\times$ Muslim pop.\ share &               &               &      -1.855   &               \\
            &               &               &     (1.335)   &               \\
\quad $\times$ Muslims $<$25\% &               &               &               &      -0.017   \\
            &               &               &               &     (0.055)   \\
\quad $\times$ Muslims 25-75\% &               &               &               &       1.257   \\
            &               &               &               &     (1.436)   \\
\quad $\times$ Muslims $>$75\% &               &               &               &      -2.845** \\
            &               &               &               &     (1.216)   \\
\hline
Country fixed effects (FE) & Yes & No & No & No \\
Ramadan year FE & Yes & No & No & No \\
District FE & No & Yes & Yes & Yes \\
Country-Ramadan year FE & No & Yes & Yes & Yes \\
Observations & 1,874,642 & 1,874,642 & 1,874,642 & 1,874,642 \\ \hline\hline
\end{tabular}
\parbox{\textwidth}{\footnotesize \emph{Notes:} Panel fixed effects regressions. Dependent variables are the extensive margins for terrorist events in panel A and terror deaths in panel B. Unlike in all other tables, we include 22,168 ambiguous events, which START classifies with high probability but not certainty as an act of terrorism. Section \ref{sec:data} introduces the time unit ``Ramadan year'' and all the variables used. Standard errors are clustered at the country-level and reported in parentheses. ***/**/* indicate statistical significance at the 1\%/5\%/10\%-level.}
\end{table}

\newpage%%%%%%%%%%%%%%%%%%%%%%%%%%%%%%%%%%%%%%%%%%%%%%%%%%%%%%

\begin{table}[h!]  
\caption*{Table C.10: Muslim countries in different regions of the world.} \label{tab:regions}\centering
\begin{tabular}{lcccc}\hline\hline
& Africa & \multicolumn{3}{c}{Asia} \\
\cline{3-5}&& All & Arab & Non-Arab \\
 & (1)   & (2)   & (3)   & (4) \\\hline
\multicolumn{5}{c}{A: Terrorist event (extensive margin)} \\\hline
Ramadan daylight hours&      -4.003***&      -1.634***&      -4.212*  &      -1.283*  \\
            &     (0.508)   &     (0.634)   &     (2.150)   &     (0.658)   \\
\hline
\multicolumn{5}{c}{B: Terror deaths (extensive margin)} \\\hline
Ramadan daylight hours&      -3.258***&      -1.602***&      -4.303** &      -1.234** \\
            &     (0.480)   &     (0.573)   &     (2.037)   &     (0.589)   \\
\hline
District FE & Yes & Yes & Yes & Yes \\
Country-Ramadan year FE & Yes & Yes & Yes & Yes \\
Observations & 100,580 &  87,514 & 28,059 & 59,455 \\ \hline\hline
\end{tabular}
\parbox{\textwidth}{\footnotesize \emph{Notes:} Panel fixed effects regressions. Dependent variables are the extensive margins for terrorist events in panel A and terror deaths in panel B. Section \ref{sec:data} introduces the time unit ``Ramadan year'' and all the variables used. The samples in columns (1) and (2) include all African countries with a Muslim population share above 75\% and all Asian countries with a Muslim population share above 75\%, respectively. Columns (3) and (4) further split the sample of these Asian countries into those that are member states of the Arab League and those that are not. Standard errors are clustered at the district-level and reported in parentheses. ***/**/* indicate statistical significance at the 1\%/5\%/10\%-level.}
\end{table}

\newpage%%%%%%%%%%%%%%%%%%%%%%%%%%%%%%%%%%%%%%%%%%%%%%%%%%%%%%

\begin{landscape}
\begin{table}[h!]
\caption*{Table C.11: Dropping individual countries} \label{tab:drop}\centering
%\begin{small}
\begin{tabular}{lccccccc}\hline\hline
 & \multicolumn{7}{c}{Omitted countries:} \\
 & Afghanistan & India & Iraq & Israel and & Nigeria & Pakistan & United \\
 & & & & Palestine & & & States \\
\cline{2-8}  
 & (1)   & (2)   & (3) & (4)   & (5) & (6)   & (7)  \\
\hline
\multicolumn{8}{c}{A: Terrorist event (extensive margin)} \\\hline
Ramadan daylight hours  & & & & \\
\quad $\times$ Muslims $<$25\%  &      -0.049   &      -0.037   &      -0.049   &      -0.049   &      -0.049   &      -0.049   &      -0.047   \\
            &     (0.075)   &     (0.071)   &     (0.075)   &     (0.075)   &     (0.075)   &     (0.075)   &     (0.084)   \\
\quad $\times$ Muslims 25-75\%  &       1.026   &       1.026   &       1.026   &       1.026   &      -0.203   &       1.026   &       1.026   \\
            &     (1.208)   &     (1.208)   &     (1.208)   &     (1.208)   &     (0.566)   &     (1.208)   &     (1.208)   \\
\quad $\times$ Muslims $>$75\%  &      -2.947** &      -2.704** &      -2.733** &      -2.721** &      -2.704** &      -2.883** &      -2.704** \\
            &     (1.213)   &     (1.209)   &     (1.237)   &     (1.205)   &     (1.209)   &     (1.180)   &     (1.209)   \\
\hline
\multicolumn{8}{c}{B: Terror deaths (extensive margin)} \\\hline
Ramadan daylight hours  & & & & \\
\quad $\times$ Muslims $<$25\%  &      -0.018   &      -0.009   &      -0.018   &      -0.017   &      -0.018   &      -0.018   &      -0.018   \\
            &     (0.040)   &     (0.037)   &     (0.040)   &     (0.040)   &     (0.040)   &     (0.040)   &     (0.045)   \\
\quad $\times$ Muslims 25-75\%  &       1.231   &       1.231   &       1.231   &       1.231   &      -0.245   &       1.231   &       1.231   \\
            &     (1.389)   &     (1.389)   &     (1.389)   &     (1.389)   &     (0.517)   &     (1.389)   &     (1.389)   \\
\quad $\times$ Muslims $>$75\%  &      -2.357** &      -2.359** &      -2.363** &      -2.361** &      -2.359** &      -2.564***&      -2.359** \\
            &     (1.036)   &     (0.970)   &     (0.997)   &     (0.970)   &     (0.970)   &     (0.924)   &     (0.970)   \\
\hline
District fixed effects (FE) & Yes & Yes & Yes & Yes & Yes & Yes & Yes \\
Country-Ramadan year FE & Yes & Yes & Yes & Yes & Yes & Yes & Yes \\
Observations & 1,859,226   & 1,846,724   & 1,869,848   & 1,873,561   & 1,838,217   & 1,873,138   & 1,727,344  \\ \hline\hline
\end{tabular}
%\end{small}
\parbox{1.25\textwidth}{\footnotesize \emph{Notes:} Panel regressions with district and country-Ramadan year fixed effects. In each column, the omitted countries are indicated in the top row. Dependent variables are the extensive margins for terrorist events in panel A and terror deaths in panel B. Section \ref{sec:data} introduces the time unit ``Ramadan year'' and all the variables used. Standard errors are clustered at the country-level and reported in parentheses. ***/**/* indicate statistical significance at the 1\%/5\%/10\%-level.}
\end{table}	
\end{landscape}

%%%%%%%%%%%%%%%%%%%%%%%%%%%%%%%%%%%%%%%%%%%%%%%%%%%%%%%%%%%%%%%%%%%%%%%%%%%%%%%%%%%%%%
\clearpage
\section*{D. Effects of Ramadan daylight hours on religious practices}

\begin{table}[h!]
\centering
\caption*{Table D.1: Ramadan daylight hours and fasting in Ramadan} 
%\begin{small}
\begin{tabular}{lcccc}
\hline \hline
 & hardly ever & some days & most days & all days \\
\hline 
Ramadan daylight hours &   -0.024         &   -0.159         &   -0.078         &    0.314*** \\
          &  (0.038)         &  (0.112)         &  (0.084)         &  (0.152)         \\ \hline
Province fixed effects (FE)	& Yes & Yes	& Yes & Yes \\
Country-Ramadan year FE	& Yes & Yes	& Yes & Yes \\
Observations & 31,890 & 31,890 & 31,890 & 31,890 \\ 
\hline\hline
\end{tabular}
\parbox{\textwidth}{\footnotesize \emph{Notes:} Panel regressions with province and country-Ramadan year fixed effects. The sample is based on Pew surveys conducted by the Pew Research Center in 13 predominantly Muslim countries and territories (namely those listed in Table~\ref{tab:support}) in the years 2002--2015. 
The dependent variables are equal to 1 if the Muslim respondent gave the answer indicated at the top of each column when asked ``how often, if at all, do you fast [in Ramadan]'', and zero otherwise
Standard errors in parentheses are clustered at the level of provinces. ***/**/* indicate statistical significance at the 1\%/5\%/10\%-level.}
%\end{small}
\end{table}

\bigskip

\begin{table}[h!]
\centering
\caption*{Table D.2: Ramadan daylight hours and praying} 
%\begin{small}
\begin{tabular}{lcccc}
\hline \hline
 & hardly ever & weekly & daily & 5x per day \\
\hline 
Ramadan daylight hours &   -0.062         &   -0.208***&    0.020         &    0.251         \\
          &  (0.043)         &  (0.065)         &  (0.092)         &  (0.175)         \\ \hline
Province fixed effects (FE)	& Yes & Yes	& Yes & Yes \\
Country-Ramadan year FE	& Yes & Yes	& Yes & Yes \\
Observations & 51,390 & 51,390 & 51,390 & 51,390 \\ 
\hline\hline
\end{tabular}
\parbox{\textwidth}{\footnotesize \emph{Notes:} Panel regressions with province and country-Ramadan year fixed effects. The sample is based on Pew surveys conducted by the Pew Research Center in 12 predominantly Muslim countries and territories (namely those listed in Table~\ref{tab:support}, except Morocco) in the years 2002--2015. 
The dependent variables are equal to 1 if the Muslim respondent gave the answer indicated at the top of each column when asked ``how often, if at all, do you pray'', and zero otherwise
Standard errors in parentheses are clustered at the level of provinces. ***/**/* indicate statistical significance at the 1\%/5\%/10\%-level.}
%\end{small}
\end{table}

%%%%%%%%%%%%%%%%%%%%%%%%%%%%%%%%%%%%%%%%%%%%%%%%%%%%%%%%%%%%%%%%%%%%%%%%%%%%%%%%%%%%%%
\newpage
\section*{E. Summary statistics and results for quarterly panel}

\begin{table}[h] \centering
\caption*{Table E.1: Summary statistics for quarterly panel, by country group} \label{tab:des_quarters}
\begin{tabular}{lcccccc} \hline\hline
 & \multicolumn{6}{c}{Muslim population share:} \\
 & \multicolumn{2}{c}{$<$ 25\%} & \multicolumn{2}{c}{25--75\%} & \multicolumn{2}{c}{$>$75\%} \\
 & Mean  & Std.Dev. & Mean  & Std.Dev. & Mean  & Std.Dev. \\
\cline{2-7} & (1)   & (2)   & (3)   & (4)   & (5)   & (6) \\\hline
Terrorist events (ext.\ margin) & 0.48  & 6.91  & 0.65  & 8.02  & 1.83  & 13.41 \\
Terror deaths (ext.\ margin) & 0.22  & 4.73  & 0.40  & 6.34  & 1.30  & 11.33 \\\hline
Observations &  \multicolumn{2}{c}{6,365,116} &  \multicolumn{2}{c}{374,684} & \multicolumn{2}{c}{758,768} \\\hline\hline
\end{tabular}
\parbox{\textwidth}{\footnotesize \emph{Notes:} Section \ref{sec:dis_short} describes the construction of the quarterly panel and Section \ref{sec:data.gtd} the terrorism data.}
\end{table} 

\begin{table} 
\caption*{Table E.2: Effect of longer Ramadan fasting on terrorism across quarters of Ramadan years} \label{tab:quarters}\centering
\begin{small}
\begin{tabular}{lcccccc}\hline\hline
  & (1)   & (2)   & (3)    & (4)   & (5)   & (6)  \\
\hline 
 & \multicolumn{3}{c}{Terrorist event (ext.\ margin)} & \multicolumn{3}{c}{Terror deaths (ext.\ margin)}\\
\hline 
Ramadan daylight hours \\
\quad $\times$ Muslims $<$25\%  &      -0.022   &      -0.024   &      -0.039   &      -0.009   &      -0.010   &      -0.025   \\
            &     (0.036)   &     (0.036)   &     (0.028)   &     (0.019)   &     (0.019)   &     (0.015)   \\
\quad \quad $\times$ Quarter 2	&               &       0.001   &       0.014   &               &      -0.001   &       0.013   \\
            &               &     (0.001)   &     (0.013)   &               &     (0.001)   &     (0.008)   \\
\quad \quad $\times$ Quarter 3	&               &       0.003** &       0.032   &               &       0.001   &       0.027   \\
            &               &     (0.001)   &     (0.037)   &               &     (0.001)   &     (0.027)   \\
\quad \quad $\times$ Quarter 4  &               &       0.003*  &       0.021   &               &       0.001   &       0.024*  \\
            &               &     (0.002)   &     (0.019)   &               &     (0.001)   &     (0.013)   \\
            
\quad $\times$ Muslims 25-75\%  &       0.512   &       0.506   &       0.522   &       0.595   &       0.593   &       0.675   \\
            &     (0.609)   &     (0.607)   &     (0.740)   &     (0.655)   &     (0.654)   &     (0.774)   \\
\quad \quad $\times$ Quarter 2	&               &       0.001   &       0.066   &               &       0.000   &       0.032   \\
            &               &     (0.001)   &     (0.098)   &               &     (0.001)   &     (0.140)   \\
\quad \quad $\times$ Quarter 3	&               &       0.011***&       0.091   &               &       0.004*  &      -0.154   \\
            &               &     (0.004)   &     (0.151)   &               &     (0.002)   &     (0.160)   \\
\quad \quad $\times$ Quarter 4  &               &       0.013***&      -0.195   &               &       0.006*  &      -0.197   \\
            &               &     (0.005)   &     (0.411)   &               &     (0.003)   &     (0.348)   \\

\quad $\times$ Muslims $>$75\%  &      -1.161** &      -1.165** &      -1.260** &      -0.917*  &      -0.919*  &      -0.945*  \\
            &     (0.531)   &     (0.530)   &     (0.487)   &     (0.504)   &     (0.503)   &     (0.491)   \\
\quad \quad $\times$ Quarter 2	&               &      -0.012   &       0.267   &               &      -0.008   &       0.022   \\
            &               &     (0.008)   &     (0.352)   &               &     (0.006)   &     (0.244)   \\
\quad \quad $\times$ Quarter 3	&               &       0.008   &       0.057   &               &       0.004   &       0.022   \\
            &               &     (0.006)   &     (0.291)   &               &     (0.005)   &     (0.256)   \\
\quad \quad $\times$ Quarter 4  &               &       0.020*  &       0.071   &               &       0.015   &       0.069   \\
            &               &     (0.011)   &     (0.218)   &               &     (0.009)   &     (0.135)   \\
\hline
District fixed effects (FE) & Yes & Yes & Yes  & Yes & Yes & Yes \\
Country-RY FE & Yes & Yes & No & Yes & Yes & No \\
Country-RY-quarter FE & No & No & Yes & No & No & Yes \\
Observations & 7,498,568 & 7,498,568 & 7,498,568 & 7,498,568 & 7,498,568 & 7,498,568 \\ 
\hline\hline
\end{tabular}
\end{small}
\parbox{\textwidth}{\footnotesize \emph{Notes:} Panel regressions with district fixed effects and either country-Ramadan year (RY) or country-Ramadan year-quarter fixed effects. Sample is based on the panel dataset with quarterly frequency explained in Section \ref{sec:dis_short}. Dependent variables the extensive margins for terrorist events in panel A and terror deaths in panel B. Section \ref{sec:data} introduces the time unit ``Ramadan year'' and the data. Standard errors are clustered at the country-level and reported in parentheses. ***/**/* indicate statistical significance at the 1\%/5\%/10\%-level.}
\end{table}

\end{document}